\magnification=1200
\noindent
\null
\centerline {\bf Damped quantum harmonic oscillator}
\vskip 1truecm
\centerline{A. Isar, ~A. Sandulescu}
\centerline{\it Department of Theoretical Physics,
Institute of Atomic Physics}
\centerline{\it POB MG-6,
Bucharest-Magurele, Romania} \vskip 1truecm

\centerline{ABSTRACT}

In the framework of the Lindblad theory for open quantum systems the damping
of the harmonic oscillator is studied. A generalization of the fundamental
constraints on quantum mechanical diffusion coefficients which appear in the
master equation for the damped quantum oscillator is presented; the
Schr\"odinger and Heisenberg representations of the Lindblad equation are
given explicitly. On the basis of these representations it is shown that
various master equations for the damped quantum oscillator used in the
literature are particular cases of the
Lindblad equation and that the majority of these equations are not satisfying
the constraints on quantum mechanical diffusion coefficients. Analytical
expressions for the first two moments of coordinate and momentum are also
obtained by using the characteristic function of the Lindblad master equation.
The master equation is transformed into Fokker-Planck equations for
quasiprobability distributions. A comparative study is made for the Glauber
$P$ representation, the antinormal ordering $Q$ representation and the Wigner
$W$ representation. It is proven that the variances for the damped harmonic
oscillator found with these representations are the same. By solving the
Fokker-Planck equations in the steady state, it is shown that the
quasiprobability distributions are two-dimensional Gaussians with widths
determined by the diffusion coefficients.
The density matrix is represented via a generating function, which is obtained
by solving a time-dependent linear partial differential equation derived from
the master equation. Illustrative examples for specific initial conditions of
the density matrix are provided.

\vskip 1truecm

{\bf 1. Introduction}

In the last two decades, more and more interest arose about the problem
of dissipation in quantum mechanics, i.e. the consistent description
of open quantum systems [1-4].The quantum description of dissipation is
important in many different areas of physics. In quantum optics we mention
the quantum theory of lasers and photon detection. There are some
directions in the theory of atomic nucleus in which dissipative processes
play a basic role. In this sense we mention the nuclear fission, giant
resonances and deep inelastic collisions of heavy ions. Dissipative
processes often occur also in many body or field-theoretical systems.

The irreversible, dissipative behaviour of the vast majority of physical
phenomena comes into an evident contradiction with the reversible nature
of our basic models. The very restrictive principles of conservative and
isolated systems are unable to deal with more complicated situations
which are determined by the features of open systems.

The fundamental quantum dynamical laws are of the reversible type. The
dynamics of a closed system is governed by the Hamiltonian that represents
 its total energy and which is a constant of motion. In this way
the paradox of irreversibility arises: the reversibility of microscopic
dynamics contrasting with the irreversibility of the macroscopic behaviour we
are trying to deduce from it.

One way to solve this paradox of irreversibility is to use models
to which Hamiltonian dynamics and Liouville's  theorem do not apply, but
the irreversible behaviour is clearly present even in the microscopic
dynamical description. The reason for replacing Hamiltonian dynamics
and Liouville's theorem is that no system is truly isolated, being subject
to uncontrollable random influences from outside. For this reason these
models are called open systems. There are two ways of treating
quantitatively their interaction with the outside. One is to introduce
specific stochastic assumptions to simulate this interaction, the other is
to treat them according to the usual laws of dynamics, by regarding the open
system as a subsystem of a larger system which is closed (i.e. which
obeys the usual laws of dynamics, with a well-defined Hamiltonian).
The dissipation arises in general from the subsystem interactions
with this larger system, often reffered to as the reservoir or bath.
The first of these two approaches has been used for the study of steady-state
transport processes, in systems obeying classical mechanics. The second of
the two approaches has been mainly used in quantum mechanics. The main
general result [1, 5-7] is that under certain conditions the time evolution
of an open system can be described by a dynamical semigroup $\Phi_t$($t\ge0$).
For a closed finite system with Liouville operator the evolution operator
is not restricted to nonnegative $t$. The importance of the dynamical
semigroup concept is that it generalizes the evolution operator to open
systems, for which there is no proper Liouville operator and no $\Phi_t$
for negative $t$. The mathematical theory of dynamical semigroups has
been developed in [1, 8-12].

The quantum mechanics of the unidimensional damped harmonic oscillator
represents a fundamental theoretical problem with applications in
different domains of quantum optics, solid state physics, molecular and
nuclear physics. In the present paper the quantum harmonic oscillator
is treated in the Lindblad axiomatic formalism of quantum dynamical
semigroups.

In Sect.2 the notion of  the quantum dynamical semigroup is
defined using the concept of a completely positive map [10]. The Lindblad
formalism replaces the dynamical group uniquely determined by its generator,
which is the Hamiltonian operator of the system, by the completely positive
dynamical semigroup with bounded generators. Then the general form of
Markovian quantum mechanical master equation is given.

In Sect.3 we give  the fundamental constraints on
quantum mechanical diffusion coefficients which appear in the corresponding
master equations [17]. The Schr\"odinger and Heisenberg representations of the
Lindblad equation are given explicitly. On the basis of these representations
it is shown that various master equations for the damped quantum oscillator
used in the literature for the description of the damped collective modes in
deep inelastic collisions or in quantum optics are particular cases of the
Lindblad equation and that the majority of these equations are not satisfying
the constraints on quantum mechanical diffusion coefficients. Explicit
expressions of the mean values and variances are also given [17,18].

In Sect.4 we solve the master equation with the characteristic function
[19]. This function is found as a solution of a corresponding partial
differential equation. By this method one can derive explicit formulae for the
centroids and variances and, in general, for moments of any order.

In Sect.5 we explore the applicability of quasiprobability distributions
to the Lindblad theory [22]. The methods of quasiprobabilities have provided
technical tools of great power for the statistical description of microscopic
systems formulated in terms of the density operator [58]. The first
quasiprobability
distribution was the one introduced by Wigner [43] in a quantum-mechanical
context. In quantum optics the $P$ representation, introduced by Glauber
[44,45,50], provided many practical applications of quasiprobabilities. The
development of quantum-mechanical master equations was combined with the
Glauber $P$ representation to give a Fokker-Planck equation for the laser
 [47,48]. The master equation of the one-dimensional damped harmonic oscillator
is transformed into Fokker-Planck equations for the Glauber $P$, antinormal
$Q$ and Wigner $W$ quasiprobability distributions associated with the density
operator. The resulting equations are solved by standard methods and
observables
directly calculated as correlations of these distribution functions. We solve
also the Fokker-Planck equations for the steady state and show that variances
found from the $P,Q$ and $W$ distributions are the same [22].

In Sect.6 we study the time evolution of the density matrix that follows
from the master equation of the damped harmonic oscillator [20,21]. We
calculate the physically relevant solutions of the master equation by applying
the method of generating function. This means that we represent the density
matrix with a generating function which is the solution of a time-dependent
partial differential equation of second order, derived from the master equation
of the damped harmonic oscillator. We discuss stationary solutions of the
generating function and derive the Bose-Einstein density matrix as example.
Then, formulas for the time development of the density matrix are presented
and illustrative examples for specific initial conditions provided. The same
method of generating function was already used by Jang [39] who studied the
damping of a collective degree of freedom coupled to a Bosonic reservoir at
finite temperature with a second order RPA master equation in the collective
subspace.

The conclusions are given in Sect.7.

\vskip 1truecm

{\bf 2. Lindblad theory for open quantum systems}

The standard quantum mechanics is Hamiltonian. The time evolution of a closed
physical system is given by a dynamical group $U_t$ which is uniquely
determined by its generator $H$, which is the Hamiltonian operator of the
system. The action of the dynamical group $U_t$ on any density matrix $\rho$
from the set $\bf D(H)$ of all density matrices of the quantum system, whose
corresponding Hilbert space is $\bf H$, is defined by
$$\rho(t)=U_t(\rho)=e^{-{i\over \hbar}Ht}\rho e^{{i\over \hbar}Ht}$$
for all $t\in(-\infty,\infty)$. We remind that, according to von Neumann,
density operators $\rho\in\bf D(H)$ are trace class ($Tr\rho<\infty$),
self-adjoint ($\rho^+=\rho$), positive ($\rho>0$) operators with
$Tr\rho=1$. All these properties are conserved by the time evolution defined
by $U_t$.

In the case of open quantum systems the main dificulty consists of finding
such time evolutions $\Phi_t$ for density operators $\rho(t)=\Phi_t(\rho)$
which preserve these von Neumann conditions for all times. From this
requirement it follows that $\Phi_t$ must have the following properties:

$(i)~  \Phi_{t}(\lambda_{1}\rho_{1}+\lambda_{2}\rho_{2})=\lambda_{1}
\Phi_{t}(\rho_{1})+\lambda_{2}\Phi_{t}(\rho_{2}); \lambda_{1},\lambda_{2}
\ge 0$ with $\lambda_{1}+\lambda_2=1,$

$(ii)~  \Phi_{t}(\rho^+)=\Phi_{t}(\rho)^+,$

$(iii)~  \Phi_{t}(\rho)>0,$

$(iv)~  Tr\Phi_{t}(\rho)=1.$

But these conditions are not restrictive enough in order to give a complete
description of the mappings $\Phi_{t}$ as in the case of the time evolutions
$U_{t}$ for closed systems. Even in the last case one has to impose other
restrictions to $U_t$, namely, it must be a group $U_{t+s}=U_tU_s.$ Also,
it is evident that in this case $U_{0}(\rho)=\rho$ and $U_{t}(\rho)\to \rho$
in the trace norm when $t\to 0$. For the dual group $\widetilde U_{t}$ acting
on the
observables $A\in \bf {B(H)},$ i.e. on the bounded operators on $\bf H$,

$$\widetilde U_{t}(A)=e^{{i\over \hbar}Ht}Ae^{-{i\over \hbar}Ht}.$$
Then $\widetilde U_{t}(AB)=\widetilde U_{t}(A)\widetilde U_{t}(B)$ and
$\widetilde U_{t}(I)=I$, where $I$ denotes the identity operator on $\bf H.$
Also, $\widetilde U_{t}(A)\to A$
ultraweakely when $t\to 0$ and $\widetilde U_{t}$ is an ultraweakely continuous
mapping [1,7,9,10,13]. These mappings have a strong positivity property
called complete positivity:
$$\sum_{i,j} B_{i}^+\widetilde U_{t}(A_{i}^+ A_{j})B_{j} \ge 0, A_{i},
B_{i}\in \bf {B(H)}.$$

Because the detailed physically plausible conditions on the systems, which
correspond to these properties are not known, it is much more convenient to
adopt an axiomatic point of view which is based mainly on the simplicity and
the succes of physical applications. Accordingly [1,7,9,10,13] it is convenient
to suppose that the time evolutions $\Phi _{t}$ for open systems are not very
different from the time evolutions for closed systems. The simplest dynamics
$\Phi _{t}$ which introduces a preferred direction in time, which is
characteristic for dissipative processes, is that in which the group condition
is replaced by the semigroup condition [6,7,11,12,14]
$$\Phi_{t+s}=\Phi_{t}\Phi_{s},~ t,s\ge 0.$$
The duality condition
$$Tr(\Phi_{t}(\rho)A)=Tr(\rho \widetilde \Phi_{t}(A)) \eqno(2.1)$$
defines $\widetilde \Phi_{t}$, the dual of $\Phi_{t}$ acting on $\bf B(H).$
Then the conditions $$Tr\Phi_{t}(\rho)=1$$and
$$\widetilde \Phi_{t}(I)=I \eqno (2.2)$$
are equivalent. Also the conditions
$$\widetilde \Phi_{t}(A)\to A \eqno (2.3)$$
ultraweakely when $t \to0$ and $$\Phi_{t}(\rho)\to \rho$$
in the trace norm when $t \to 0,$ are equivalent.
For the semigroups with the properties (2.2), (2.3) and
$$A\ge 0 \to \widetilde \Phi_{t}(A) \ge 0,$$
it is well known that there exists a (generally bounded) mapping
$\widetilde L$-the generator of $\widetilde \Phi_{t}.$
$\widetilde \Phi_{t}$ is uniquely determined by $\widetilde L.$
The dual generator of the dual semigroup $\Phi_{t}$ is denoted by $L$:
$$Tr(L(\rho)A)=Tr(\rho \widetilde L(A)).$$
The evolution equations by which $L(\widetilde L)$ determine uniquely
$\Phi_{t}(\widetilde \Phi_{t})$ are
$${d\Phi_{t}(\rho)\over dt}=L(\Phi_{t}(\rho)) \eqno (2.4)$$
and
$${d\widetilde \Phi_{t}(A)] \over dt}=\widetilde L(\widetilde \Phi_{t}(A)),
\eqno (2.5)$$
respectively, in the Schr\"odinger and Heisenberg picture.
These equations replace in the case of open systems the von-Neumann-Liouville
equations
$${dU_{t}(\rho)\over dt}=-{i\over \hbar}[H,U_{t}(\rho)]$$
and
$${d\widetilde U_{t}(A)\over dt}={i\over \hbar}[H,\widetilde U_{t}(A)],$$
respectively.

For any applications eqs.(2.4) and (2.5) are only useful if the detailed
structure of the generator $L(\widetilde L)$ is known and can be related to
the concrete properties of the open systems, which are described by such
equations.

Such a structural theorem was obtained by Lindblad [10] for the class of
dynamical semigroups $\widetilde \Phi_{t}$ which are completely positive and
norm continuous. For such semigroups the generator $\widetilde L$ is bounded.
In many applications the generator is unbounded.

A bounded mapping, $\widetilde L:\bf {B(H)}\to \bf {B(H)}$ which satisfies
$\widetilde L(I)=0, ~\widetilde L(A^+)=\widetilde L(A)^+$ and
$$\widetilde L(A^+ A)-\widetilde L(A^+)A-A^+ \widetilde L(A)\ge 0$$
is called dissipative. The 2-positivity property of the completely positive
mapping $\widetilde \Phi_{t}$:
$$\widetilde \Phi_{t}(A^+ A)\ge \widetilde \Phi_{t}(A^+)\widetilde
 \Phi_{t}(A),$$
with equality at $t=0$, implies that $\widetilde L$ is dissipative. Lindblad
[10] has shown that conversely, the dissipativity of $\widetilde L$ implies
that $\widetilde \Phi_{t}$ is 2-positive. $\widetilde L$ is called completely
dissipative if all trivial extensions of $\widetilde L$ are
 dissipative.
Lindblad has also shown that there exists a one-to-one correspondence between
the completely positive norm continuous semigroups $\widetilde \Phi_{t}$ and
completely dissipative generators $\widetilde L$. The structural theorem of
Lindblad gives the most general form of a completely dissipative mapping
$\widetilde L$ [10,11]:

{\it Theorem}: $\widetilde L$ is completely dissipative and ultraweakely
continuous if and only if it is of the form
$$\widetilde L(A)={i\over \hbar}[H,A]+{1\over 2\hbar}
\sum_{j} (V_{j}^+[A,V_{j}]+[V_{j}^+,A]V_{j}), \eqno (2.6)$$
where $V_{j}, \sum_{j} V_{j}^+ V_{j}\in{\bf B(H)}, ~H\in {\bf B(H)}_{s.a.}$.

The dual generator on the state space (Schr\"odinger picture) is of the form
$$L(\rho)=-{i\over\hbar}[H,\rho]+{1\over 2\hbar}\sum_{j}([V_{j}\rho,V_{j}^+]
+[V_{j},\rho V_{j}^+]). \eqno (2.7)$$
Eqs.(2.4) and (2.7) give an explicit form for the most general
time-homogeneous quantum mechanical Markovian master equation with a bounded
Liouville operator.

Talkner [16] has shown that the assumption of a semigroup dynamics is only
applicable in the limit of weak coupling of the subsystem with its
environment, i.e. for long relaxation times.

We should like to mention that all Markovian master equations found in the
literature are of this form after some rearrangement of terms, even for
unbounded generators.

It is also an empirical fact that for many physically interesting situations
the time evolutions $\Phi_{t}$ drive the system toward a unique final state
$\rho (\infty)=\lim_{t\to \infty} \Phi_{t}(\rho(0))$ for all
$\rho (0)\in \bf D(H)$.

The evolution equations of Lindblad, being operator equations, the problem of
finding their solutions is, in general, rather difficult. In cases when the
equations are exactly solvable, these solutions give complete informations
about the studied problem - they permit the calculation of expectation values
of the observables at any moment.

\vskip 1truecm

{\bf 3. Master equations for damped quantum harmonic oscillator}

In this Section the case of damped quantum harmonic oscillator is considered
in the spirit of the ideas presented in the previous Section. The basic
assumption is that the general form (2.7) of a bounded mapping $L$ given by
Lindblad theorem [10] is also valid for an unbounded completely dissipative
mapping $L$:
$$L(\rho)=-{i\over \hbar}[H,\rho]+{1\over 2\hbar}\sum_{j} ([V_{j}\rho,V_{j}
^+]+[V_{j},\rho V_{j}^+]). \eqno (3.1)$$
This assumption gives one of the simplest way to construct an appropriate
model for this quantum dissipative system. Another simple condition imposed
to the operators $H,V_{j},V_{j}^+$ is that they are functions of the basic
observables of the one-dimensional quantum mechanical system $q$ and $p$
with $[q,p]=i\hbar I$, where $I$ is the identity operator on $\bf H$ of such
kind that the obtained model is exactly solvable. A precise version for this
last condition is that linear spaces spanned by the first degree (respectively
second degree) noncommutative polynomials in $p$ and $q$ are invariant to the
action of the completely dissipative mapping $L$. This condition implies [15]
that $V_{j}$ are at most the first degree polynomials in $p$ and $q$ and $H$
is at most a second degree polynomial in $p$ and $q$.

Beacause in the linear space of the first degree polynomials in $p$ and $q$
the operators $p$ and $q$ give a basis, there exist only two $C$-linear
independent operators $V_{1},V_{2}$ which can be written in the form
$$V_{i}=a_{i}p+b_{i}q,~i=1,2$$
with $a_{i},b_{i}=1,2$ complex numbers [15]. The constant term is omitted
because its contribution to the generator $L$ is equivalent to terms in $H$
linear in $p$ and $q$ which for simplicity are chosen to be zero. Then $H$ is
chosen of the form
$$H=H_{0}+{\mu \over 2}(pq+qp),~~~H_{0}={1\over 2m}p^2+{m\omega^2
\over 2}q^2. \eqno (3.2)$$
With these choices the Markovian master equation can be written:
$${d\rho \over dt}=-{i\over \hbar}[H_{0},\rho]-{i\over 2\hbar}(\lambda +\mu)
[q,\rho p+p\rho]+{i\over 2\hbar}(\lambda -\mu)[p,\rho q+q\rho]-$$
$$-{D_{pp}\over {\hbar}^2}[q,[q,\rho]]-{D_{qq}\over {\hbar}^2}[p,[p,\rho]]+
{D_{pq}\over {\hbar}^2}([q,[p,\rho]]+[p,[q,\rho]]). \eqno (3.3)$$
Here we used the notations:
$$D_{qq}={\hbar\over 2}\sum_{j=1,2}{\vert a_{j}\vert}^2,
  D_{pp}={\hbar\over 2}\sum_{j=1,2}{\vert b_{j}\vert}^2,
D_{pq}=D_{qp}=-{\hbar\over 2}Re\sum_{j=1,2}a_{j}^*b_{j},
\lambda=-Im\sum_{j=1,2}a_{j}^*b_{j},$$
where $D_{pp},D_{qq}$ and $D_{pq}$ are the diffusion coefficients and
$\lambda$ the friction constant. They satisfy the following fundamental
constraints as shown in [17]:

$i)~D_{pp}>0$

$ii)~D_{qq}>0 ~~~~~~~~~~~~~~~~~~~~~~~~~~~~~~~~~~~~~~~~~~~~~~~~~~~~~~~~~~ (3.4)$

$iii)~D_{pp}D_{qq}-{D_{pq}}^2\ge {\lambda}^2{\hbar}^2/4.$

Introducing the annihilation and creation operators
$$a={1\over \sqrt{2\hbar}}(\sqrt{m\omega}q+{i\over \sqrt{m\omega}}p),$$
$$a^+={1\over\sqrt{2\hbar}}(\sqrt{m\omega}q-{i\over\sqrt{m\omega}}p),\eqno
(3.5)$$
obeying the commutation relation $[a,a^+]=1$, we have
$$H_{0}=\hbar\omega(a^+a+{1\over 2}) \eqno (3.6)$$
and the master equation has the form
$${d\rho\over dt}={1\over 2}(D_{1}-\mu)(\rho a^+ a^+-a^+\rho a^+)+
{1\over 2}(D_{1}+\mu)(a^+ a^+\rho-a^+\rho a^+)+$$
$$+{1\over 2}(D_{2}-\lambda-i\omega)(a^+\rho a-\rho aa^+)+
{1\over2}(D_{2}+\lambda+i\omega)(a\rho a^+-a^+ a\rho)+h.c.,\eqno (3.7)$$
where
$$D_{1}={1\over\hbar}(m\omega D_{qq}-{D_{pp}\over m\omega}+2iD_{pq}),$$
$$D_{2}={1\over\hbar}(m\omega D_{qq}+{D_{pp}\over m\omega}). \eqno (3.8)$$

In the literature, equations of this kind are encountered in concrete
theoretical models for the description of different physical phenomena in
quantum optics, the damping of collective modes in deep
inelastic collisions of heavy ions or in the quantum mechanical
description of the dissipation for the one-dimensional harmonic
oscillator. In the following we show that all these master
equations are particular cases of the Lindblad equation and that the
majority of these equations are not satisfying the constraints on quantum
mechanical diffusion coefficients, and therefore the uncertainty principle
is violated.

1) The Dekker master equation for the damped quantum harmonic oscillator
[4,23-26] supplemented with the fundamental constraints (3.4) obtained in
[23] from the condition that the time evolution of this master equation does
not violate the uncertainty principle at any time, is a particular case of
the Lindblad master equation (3.7) when $\mu=\lambda$.

2) The quantum master equation considered in [27,28] by Hofmann et al. for
treating the charge equilibration process as a collective high frequency mode
is a particular case of the Lindblad master equation (3.3) if $\lambda=
\gamma(\omega)/2m=\mu, D_{qq}=0, D_{pp}=\gamma(\omega)T^*(\omega),
D_{pq}=0,$ but the fundamental constraints (3.4) are not satisfied.

3) For the quantum master equation considered in [29] for the description
of heavy ion collisions we have $\lambda=\mu=\gamma/2, D_{pp}=D, D_{qq}=0,
D_{pq}=D_{qp}=-d/2$ and consequently the fundamental constraints are not
fulfilled.

4) In [30], Spina and Weidenm\"uller considered two kinds of master equations
I and II for describing the damping of collective modes in deep inelastic
collisions of heavy ions. Eq.I can be obtained from eq.(3.3) by replacing
$H_{0}$ with $H_{0}-{1\over 2}Am\omega q^2+f(t)q$ and setting
$\lambda=\mu=\Gamma/2, D_{pp}=D/2, D_{qq}=0$ and
$D_{pq}=D_{qp}=B/2$. Then the constraints (3.4) are not satisfied.
Eq.II is obtained from (3.3) by putting $H_{0}-(1/2)A^{II}m\omega q^2-
(1/2m\omega)A^{II}p^2+f(t)q$ and $\Gamma_{R}^{II}=\Gamma_{p}^{II}=
\lambda, \mu=0, D_{pp}=D_{p}^{II}/2, D_{qq}=D_{R}^{II}/2,
D_{pq}=0$ and the last condition (3.4) is satisfied for all values of the
parameters.

5) The master equation for the density operator of the electromagnetic field
mode coupled to a squeezed bath [31,32] can be obtained from the master
equation (3.7) if we set
$$\mu=0, \lambda=\gamma, {1\over 2\lambda}(\lambda-
{m\omega D_{qq}\over \hbar}-{D_{pp}\over \hbar m\omega})=-N, {1\over 2\lambda}
({m\omega D_{qq}\over \hbar}-{D_{pp}\over \hbar m\omega}+2i{D_{pq}\over \hbar})
=M.$$

6) The master equation for the density operator of  a harmonic oscillator
coupled to an environment of harmonic oscillators considered in [33-36] is a
particular case of the master equation (3.7) if we put
$$\lambda=\mu=\gamma,D_{qq}=0,D_{pq}=0,{1\over 2\lambda}(\lambda-{D_{pp}\over
\hbar m\omega})=-\bar n$$
and the fundamental constraints (3.4) are not
fulfilled.

7) The master equation written in [37] for different models of correlated-
emission lasers can also be obtained from the master equation (3.7) by putting
$${1\over 2}(D_{1}+\mu)=\Lambda_{4},{1\over 2}(D_{1}-\mu)=\lambda_{3},
{1\over 2}(D_{2}+\lambda+i\omega)= \Lambda_{2},
{1\over 2}(D_{2}-\lambda-i\omega)=\Lambda_{1}.$$

8) Two master equations were introduced by Jang in [38,39], where the nuclear
dissipative pocess is described as the damping of a collective degree of
freedom coupled to a bosonic reservoir at finite temperature. The resulting
RPA master equation within the observed collective subspace is derived in a
purely dynamical way. The master equation written in [38] in the resonant
approximation (rotating-wave approximation) can be obtained as a particular
case of the Lindblad master equation (3.7). For this one has to set
$$D_{pp}=m^2\omega^2D_{qq}, D_{pq}=\mu=0,
{4m\omega D_{qq}\over \hbar}=(2<n>+1)\Gamma,\lambda={\Gamma\over 2},$$
where $<n>$ is the average number of the RPA collective phonons at thermal
equilibrium and $\Gamma$ is the width (friction parameter). The fundamental
constraints (3.4) are fulfilled in this case.

The master equation derived recently [39] in order to extend the calculations
carried out in [33] with the before-mentioned master equation in the resonant
approximation, can also be obtained as a particular case of the master
equation (3.7) by taking
$$D_{qq}=D_{pq}=0, D_{pp}={\hbar m\omega\over 2}(2<n>+1)\Gamma,
\mu=\lambda={\Gamma\over 2}$$
or $D_{2}=D_{1}=(2<n>+1)\Gamma/2$ and in this case the fundamental constraints
(3.4) are not fulfilled.

The following notations will be used:
$$\sigma_{q}(t)=Tr(\rho(t)q),$$
$$\sigma_{p}(t)=Tr(\rho(t)p),$$
$$\sigma_{qq}=Tr(\rho(t)q^2)-\sigma_{q}^2(t), \eqno (3.9)$$
$$\sigma_{pp}=Tr(\rho(t)p^2)-\sigma_{p}^2(t),$$
$$\sigma_{pq}(t)=Tr(\rho(t){pq+qp\over 2})-\sigma_{p}(t)\sigma_{q}(t).$$

In the Heisenberg picture the master equation has the following symmetric
form:
$${d\widetilde\Phi_{t}(A)\over dt}=\widetilde L(\widetilde\Phi_{t}(A))=
{i\over\hbar}[H_{0},\widetilde\Phi_{t}(A)]-{i\over 2\hbar}(\lambda+\mu)
([\widetilde\Phi_{t}(A),q]p+p[\widetilde\Phi_{t}(A),q])+$$
$$+{i\over 2\hbar}(\lambda-\mu)(q[\widetilde\Phi_{t}(A),p]+[\widetilde\Phi_{t}
(A),p]q)-{D_{pp}\over\hbar^2}[q,[q,\widetilde\Phi_{t}(A)]]-$$
$$-{D_{qq}\over\hbar^2}[p,[p,\widetilde\Phi_{t}(A)]]+{D_{pq}\over\hbar^2}
([p,[q,\widetilde\Phi_{t}(A)]]+[q,[p,\widetilde\Phi_{t}(A)]]).$$

Denoting by $A$ any selfadjoint operator we have
$$\sigma_{A}(t)=Tr(\rho(t)A), \sigma_{AA}(t)=Tr(\rho(t)A^2)-\sigma_{A}^2(t).$$
It follows that
$${d\sigma_{A}(t)\over dt}=Tr L(\rho(t))A=Tr\rho(t)\widetilde L(A) \eqno
(3.10)$$ and
$${d\sigma_{AA}(t)\over dt}=Tr L(\rho(t))A^2-2{d\sigma_{A}(t_)\over dt}
\sigma_{A}(t)=Tr\rho(t)\widetilde L(A^2)-2\sigma_{A}(t)Tr\rho(t)
\widetilde L(A). \eqno (3.11)$$

An important consequence of the precise version of solvability condition
formulated at the beginning of the present Section is the fact that when
$A$ is put equal to $p$ or $q$ in (3.10) and (3.11), then
$d\sigma_{p}(t)/dt$ and $d\sigma_{q}(t)/dt$ are functions only
of $\sigma_{p}(t)$ and $\sigma_{q}(t)$ and $d\sigma_{pp}(t)/dt,
d\sigma_{qq}(t)/dt$ and $d\sigma_{pq}(t)/dt$ are functions only
of $\sigma_{pp}(t),\sigma_{qq}(t)$ and $\sigma_{pq}(t)$. This fact allows an
immediate determination of the functions of time $\sigma_{p}(t),\sigma_{q}(t),
\sigma_{pp}(t),\sigma_{qq}(t),$ $\sigma_{pq}(t)$. The results are the
following:
$${d\sigma_{q}(t)\over dt}=-(\lambda-\mu)\sigma_{q}(t)+{1\over m}\sigma_{p}
(t),$$
$${d\sigma_{p}(t)\over dt}=-m\omega^2\sigma_{q}(t)-(\lambda+\mu)\sigma_{p}
(t) \eqno(3.12)$$
and
$${d\sigma_{qq}(t)\over dt}=-2(\lambda-\mu)\sigma_{qq}(t)+{2\over m}
\sigma_{pq}(t)+2D_{qq},$$
$${d\sigma_{pp}\over dt}=-2(\lambda+\mu)\sigma_{pp}(t)-2m\omega^2\sigma_{pq}(t)
+2D_{pp}, \eqno(3.13)$$
$${d\sigma_{pq}(t)\over dt}=-m\omega^2\sigma_{qq}(t)+{1\over m}\sigma_{pp}(t)
-2\lambda\sigma_{pq}(t)+2D_{pq}.$$
 All equations considered in various papers in
connection with damping of collective modes in deep inelastic collisions are
obtained as particular cases of Eqs. (3.13), as we already mentioned before.

The integration of Eqs.(3.12) is straightforward. There are two cases: {\it a)}
$\mu>\omega$ (overdamped) and {\it b)} $\mu<\omega$ (underdamped). If $S(t)$
denotes
the vector $${\sigma_q(t)\choose\sigma_p(t)}$$ and $M$ the $2\times 2$ matrix
$$M=\left(\matrix{-(\lambda-\mu)&1/m\cr
-m\omega^2&-(\lambda+\mu)\cr}\right),$$
then (3.12) becomes
$${dS(t)\over dt}=MS(t). \eqno(3.14)$$
Now $M$ can be written as $M=N^{-1}FN$ with $F$ a diagonal matrix. It follows
that the solution of (3.14) is $$S(t)=N^{-1}e^{Ft}NS(0).$$
In the case {\it a)} with the notation $\nu^2=\mu^2-\omega^2$ the matrices
$N,N^{-1}$ and $F$ are given by
$$N=\left(\matrix{m\omega^2&\mu+\nu\cr
m\omega^2&\mu-\nu\cr}\right),$$
$$N^{-1}={1\over 2m\omega^2\nu}\left(\matrix{-(\mu-\nu)&\mu+\nu\cr
m\omega^2&-m\omega^2\cr}\right),$$
$$F=\left(\matrix{-(\lambda+\nu)&0\cr
0&-(\lambda-\nu)\cr}\right).$$
Then
$$N^{-1}e^{Ft}N=e^{-\lambda t}\left(\matrix{\cosh\nu t+{\mu\over\nu}\sinh\nu t&
{1\over m\nu}\sinh\nu t\cr
-{m\omega^2\over\nu}\sinh\nu t&\cosh\nu t-{\mu\over\nu}\sinh\nu t\cr}\right),$$
i.e.,
$$\sigma_q(t)=e^{-\lambda t}((\cosh\nu t+{\mu\over\nu}\sinh\nu t)\sigma_q(0)+
{1\over m\nu}\sinh\nu t\sigma_p(0)),$$
$$\sigma_p(t)=e^{-\lambda t}(-{m\omega^2\over\nu}\sinh\nu t\sigma_q(0)+
(\cosh\nu t-{\mu\over\nu}\sinh\nu t)\sigma_p(0)).\eqno(3.15)$$
If $\lambda>\nu$, then $\sigma_q(\infty)=\sigma_p(\infty)=0$. If $\lambda<\nu$,
then $\sigma_q(\infty)=\sigma_p(\infty)=\infty$. In the case {\it b)} with
the notation $\Omega^2=\omega^2-\mu^2$, the matrices $N,N^{-1}$ and $F$ are
given by
$$N=\left(\matrix{m\omega^2&\mu+i\Omega\cr
m\omega^2&\mu-i\Omega\cr}\right),$$
$$N^{-1}={1\over 2im\omega^2\Omega}\left(\matrix{-(\mu-i\Omega)&\mu+i\Omega\cr
m\omega^2&-m\omega^2\cr}\right),$$
$$F=\left(\matrix{-(\lambda+i\Omega)&0\cr
0&-(\lambda-i\Omega)\cr}\right).$$
Then
$$N^{-1}e^{Ft}N=e^{-\lambda t}\left(\matrix{\cos\Omega t+{\mu\over\Omega}
\Omega t&{1\over m\Omega}\sin\Omega t\cr
-{m\omega^2\over\Omega}\sin\Omega t&\cos\Omega t-{\mu\over\Omega}\sin\Omega t
\cr}
\right),$$
i.e.,
$$\sigma_q(t)=e^{-\lambda t}((\cos\Omega t+{\mu\over\Omega}\sin\Omega t)
\sigma_q(0)+{1\over m\Omega}\sin\Omega t\sigma_p(0)),$$
$$\sigma_p(t)=e^{-\lambda t}(-{m\omega^2\over\Omega}\sin\Omega t\sigma_q(0)+
(\cos\Omega t-{\mu\over\Omega}\sin\Omega t)\sigma_p(0)) \eqno(3.16)$$
and $\sigma_q(\infty)=\sigma_p(\infty)=0.$

In order to integrate Eqs.(3.13), it is convenient to consider the vector
$$X(t)=\left(\matrix{m\omega\sigma_{qq}(t)\cr
{1\over m\omega}\sigma_{pp}(t)\cr
\sigma_{pq}(t)\cr}\right).$$
Then the system of equations (3.13) can be written in the form
$${dX(t)\over dt}=RX(t)+D,$$
where $R$ is the following $3\times 3$ matrix
$$R=\left(\matrix{-2(\lambda-\mu)&0&2\omega\cr
0&-2(\lambda+\mu)&-2\omega\cr
-\omega&\omega&-2\lambda\cr}\right)$$
and $D$ is the following vector
$$D=\left(\matrix{2m\omega D_{qq}\cr
{2\over m\omega}D_{pp}\cr
2D_{pq}\cr}\right).$$
Then there exists a matrix $T$ with property $T^2=I$ where $I$ is the identity
matrix and a diagonal matrix $K$ such that $R=TKT$. From this it follows that
$$X(t)=(Te^{Kt}T)X(0)+T(e^{Kt}-I)K^{-1}TD.\eqno(3.17)$$
In the overdamped case $(\mu>\omega)$ the matrices $T$ and $K$ are given by
$$T={1\over 2\nu}\left(\matrix{\mu+\nu&\mu-\nu&2\omega\cr
\mu-\nu&\mu+\nu&2\omega\cr
-\omega&-\omega&-2\mu\cr}\right)$$
and
$$K=\left(\matrix{-2(\lambda-\nu)&0&0\cr
0&-2(\lambda+\nu)&0\cr
0&0&-2\lambda\cr}\right)$$
with $\nu^2=\mu^2-\omega^2$.

In the underdamped case $(\mu<\omega)$ the matrices $T$ and $K$ are given by
$$T={1\over 2i\Omega}\left(\matrix{\mu+i\Omega&\mu-i\Omega&2\omega\cr
\mu-i\Omega&\mu+i\Omega&2\omega\cr
-\omega&-\omega&-2\mu\cr}\right)$$
and
$$K=\left(\matrix{-2(\lambda-i\Omega)&0&0\cr
0&-2(\lambda+i\Omega)&0\cr
0&0&-2\lambda\cr}\right)$$
with $\Omega^2=\omega^2-\mu^2$.

From (3.17) it follows that
$$X(\infty)=-(TK^{-1}T)D=-R^{-1}D \eqno(3.18)$$
(in the overdamped case the restriction $\lambda>\nu$ is necessary). Then
Eq.(3.17) can be written in the form
$$X(t)=(Te^{Kt}T)(X(0)-X(\infty))+X(\infty).\eqno(3.19)$$
Also
$${dX(t)\over dt}=(TKe^{Kt}T)(X(0)-X(\infty))=R(X(t)-X(\infty))$$
and
$${dX(t)\over dt}\vert_{t=0}=(TKT)(X(0)-X(\infty))=R(X(0)-X(\infty)).$$
The formula (3.18) is remarcable because it gives a very simple connection
between the asymptotic values of $\sigma_{qq}(t),\sigma_{pp}(t),\sigma_{pq}(t)$
and the diffusion coefficients $D_{qq},D_{pp},D_{pq}$. As an immediate
consequence of (3.18) this connection is {\it the same} for both cases,
underdamped and overdamped, and has the following explicit form:
$$\sigma_{qq}(\infty)={1\over 2(m\omega)^2\lambda(\lambda^2+\omega^2-\mu^2)}
((m\omega)^2(2\lambda(\lambda+\mu)+\omega^2)D_{qq}+$$
$$+\omega^2D_{pp}+2m\omega^2(\lambda+\mu)D_{pq}),$$
$$\sigma_{pp}(\infty)={1\over 2\lambda(\lambda^2+\omega^2-\mu^2)}((m\omega)^2
\omega^2D_{qq}+(2\lambda(\lambda-\mu)+\omega^2)D_{pp}-2m\omega^2(\lambda-
\mu)D_{pq}),\eqno(3.20)$$
$$\sigma_{pq}(\infty)={1\over 2m\lambda(\lambda^2+\omega^2-\mu^2)}(-(\lambda+
\mu)(m\omega)^2D_{qq}+(\lambda-\mu)D_{pp}+2m(\lambda^2-\mu^2)D_{pq}).$$
These relations show that the asymptotic values $\sigma_{qq}(\infty),
\sigma_{pp}(\infty),\sigma_{pq}(\infty)$ do not depend on the initial values
$\sigma_{qq}(0),\sigma_{pp}(0),\sigma_{pq}(0)$. In other words,
$$R^{-1}={-1\over 4\lambda(\lambda^2+\omega^2-\mu^2)}\left(\matrix{2\lambda
(\lambda+\mu)+\omega^2&\omega^2&2\omega(\lambda+\mu)\cr
\omega^2&2\lambda(\lambda-\mu)+\omega^2&-2\omega(\lambda-\mu)\cr
-(\lambda+\mu)\omega&(\lambda-\mu)\omega&2(\lambda^2-\mu^2)\cr}\right).$$
Conversely, if the relations $D=-RX(\infty)$ are considered, i.e.,
$$\left(\matrix{2m\omega D_{qq}\cr
{2\over m\omega}D_{pp}\cr
2D_{pq}\cr}\right)=-\left(\matrix{-2(\lambda-\mu)&0&2\omega\cr
0&-2(\lambda+\mu)&-2\omega\cr
-\omega&\omega&-2\lambda\cr}\right)\left(\matrix{m\omega\sigma_{qq}(\infty)
\cr
{1\over m\omega}\sigma_{pp}(\infty)\cr
\sigma_{pq}(\infty)\cr}\right),$$
then $$D_{qq}=(\lambda-\mu)\sigma_{qq}(\infty)-{1\over m}\sigma_{pq}(\infty),$$
$$D_{pp}=(\lambda+\mu)\sigma_{pp}(\infty)+m\omega^2\sigma_{pq}(\infty),\eqno
(3.21)$$
$$D_{pq}={1\over2}(m\omega^2\sigma_{qq}(\infty)-{1\over m}\sigma_{pp}(\infty)+
2\lambda\sigma_{pq}(\infty)).$$
Hence, from (3.4) the fundamental constraints on $\sigma_{qq}(\infty),
\sigma_{pp}(\infty)$ and $\sigma_{pq}(\infty)$ follow:
$$D_{qq}=(\lambda-\mu)\sigma_{qq}(\infty)-{1\over m}\sigma_{pq}(\infty)>0,$$
$$D_{pp}=(\lambda+\mu)\sigma_{pp}(\infty)+m\omega^2\sigma_{pq}(\infty)>0,$$
$$D_{qq}D_{pp}-D_{pq}^2=(\lambda^2-\mu^2)\sigma_{qq}(\infty)\sigma_{pp}(\infty)
-\omega^2\sigma_{pq}^2(\infty)+$$
$$+(\lambda-\mu)m\omega^2\sigma_{qq}(\infty)\sigma_{pq}(\infty)-{(\lambda+\mu)
\over m}\sigma_{pp}(\infty)\sigma_{pq}(\infty)-$$
$$-{1\over 4}(m\omega^2)^2\sigma_{qq}^2(\infty)-{1\over 4m^2}\sigma_{pp}^2
(\infty)-\lambda^2\sigma_{pq}^2(\infty)+{1\over 2}\omega^2\sigma_{qq}
(\infty)\sigma_{pp}(\infty)-$$
$$-m\omega^2\lambda\sigma_{qq}(\infty)\sigma_{pq}(\infty)+{\lambda\over m}
\sigma_{pp}(\infty)\sigma_{pq}(\infty)\ge{\lambda^2\hbar^2\over 4}.
\eqno(3.22)$$

The constraint (3.22) can be put in a more clear form:
$$4(\lambda^2+\omega^2-\mu^2)(\sigma_{qq}(\infty)\sigma_{pp}(\infty)-
\sigma_{pq}(\infty)^2)-$$
$$-(m\omega^2\sigma_{qq}(\infty)+{1\over m}\sigma_{pp}(\infty)+2\mu\sigma_{pq}
(\infty))^2\ge\hbar^2\lambda^2. \eqno(3.23)$$
If $\mu<\omega$ (the underdamped case), then $\lambda^2+\omega^2-\mu^2>
\lambda^2$. If $\mu>\omega$ (the overdamped case), then $0\le\lambda^2+
\omega^2-\mu^2<\lambda^2$ $(\lambda>\nu)$ and the constraint (3.23) is more
strong than the uncertainty inequality $\sigma_{qq}(\infty)\sigma_{pp}
(\infty)-\sigma_{pq}^2(\infty)\ge\hbar^2/4$. By using the fact that
the linear positive mapping ${\bf B(H)}\to C$ defined by $A\to Tr(\rho A)$
is completely positive (hence 2-positive), in [17] the following inequality
was obtained:
$$D_{qq}\sigma_{pp}(t)+D_{pp}\sigma_{qq}(t)-2D_{pq}\sigma_{pq}(t)\ge
{\hbar^2\lambda\over 2}.$$
From this inequality which must be valid for all values of $t\in(0,\infty)$
it follows that
$$D_{qq}\sigma_{pp}(\infty)+D_{pp}\sigma_{qq}(\infty)-2D_{pq}\sigma_{pq}
(\infty)\ge{\hbar^2\lambda\over 2}.$$
Using Eq. (3.21) this inequality is equivalent with the uncertainty inequality
$$\sigma_{qq}(\infty)\sigma_{pp}(\infty)-\sigma_{pq}(\infty)^2\ge{\hbar^2
\over 4}.$$
A restriction connecting the initial values $\sigma_{qq}(0),\sigma_{pp}(0),
\sigma_{pq}(0)$ with the asymptotic values $\sigma_{pp}(\infty),\sigma_{qq}(
\infty),\sigma_{pq}(\infty)$ is also obtained:
$$D_{qq}\sigma_{pp}(0)+D_{pp}\sigma_{qq}(0)-2D_{pq}\sigma_{pq}(0)\ge
{\hbar^2\lambda\over 2}.$$
More explicitly
$$\lambda(\sigma_{qq}(\infty)\sigma_{pp}(0)+\sigma_{pp}(\infty)\sigma_{qq}(0)-
2\sigma_{pq}(\infty)\sigma_{pq}(0))-$$
$$-\mu(\sigma_{qq}(\infty)\sigma_{pp}(0)-\sigma_{pp}(\infty)\sigma_{qq}(0))-
{1\over m}(\sigma_{pq}(\infty)\sigma_{pp}(0)-\sigma_{pp}(\infty)\sigma_{pq}
(0))+$$
$$+m\omega^2(\sigma_{pq}(\infty)\sigma_{qq}(0)-\sigma_{qq}(\infty)\sigma_{pq}
(0))\ge{\hbar^2\lambda\over 2}.\eqno(3.24)$$

If the asymptotic state is a Gibbs state
$$\rho_G(\infty)=e^{-{H_0\over kT}}/Tr(e^{-{H_0\over kT}}),$$
then
$$\sigma_{qq}(\infty)={\hbar\over 2m\omega}\coth{\hbar\omega\over 2kT},
 \sigma_{pp}(\infty)={\hbar m\omega\over 2}\coth{\hbar\omega\over 2kT},
 \sigma_{pq}(\infty)=0\eqno(3.25)$$
and
$$D_{pp}={\lambda+\mu\over 2}\hbar m\omega\coth{\hbar\omega\over 2kT},
 D_{qq}={\lambda-\mu\over 2}{\hbar\over m\omega}\coth{\hbar\omega\over 2kT},
 D_{pq}=0\eqno((3.26)$$
and the fundamental constraints (3.4) are satisfied only if $\lambda>\mu$ and
[15]:
$$(\lambda^2-\mu^2)(\coth{\hbar\omega\over 2kT})^2\ge\lambda^2.$$

If the initial state is the ground state of the harmonic oscillator, then
$$\sigma_{qq}(0)={\hbar\over 2m\omega}, \sigma_{pp}(0)={m\hbar\omega\over 2},
 \sigma_{pq}(0)=0.$$
Then (3.24) becomes
$$\lambda(\sigma_{qq}(\infty)m\omega+{\sigma_{pp}(\infty)\over m\omega})-
\mu(\sigma_{qq}(\infty)m\omega-{\sigma_{pp}(\infty)\over m\omega})\ge
\hbar\lambda.$$
For example, in the case (3.25), this implies $\coth{\hbar\omega/2kT}\ge1$
which is always valid.

Now, the explicit time dependence of $\sigma_{qq}(t),\sigma_{pp}(t),\sigma_{pq}
(t)$ will be given for both under- and overdamped cases. From Eq. (3.19) it
follows that in order to obtain this explicit time dependence it is necessary
to obtain the matrix elements of $Te^{Kt}T$. In the overdamped case
$(\mu>\omega),\nu^2=\mu^2-\omega^2$ we have
$$Te^{Kt}T={e^{-2\lambda t}\over 2\nu^2}\left(\matrix{a_{11}&a_{12}&a_{13}\cr
a_{21}&a_{22}&a_{23}\cr
a_{31}&a_{32}&a_{33}\cr}\right),$$
with

$a_{11}=(\mu^2+\nu^2)\cosh 2\nu t+2\mu\nu\sinh 2\nu t-\omega^2,$

$a_{12}=(\mu^2-\nu^2)\cosh 2\nu t-\omega^2,$

$a_{13}=2\omega(\mu\cosh 2\nu t+\nu\sinh 2\nu t-\mu),$

$a_{21}=(\mu^2-\nu^2)\cosh 2\nu t-\omega^2,$

$a_{22}=(\mu^2+\nu^2)\cosh 2\nu t-2\mu\nu\sinh 2\nu t-\omega^2,$~~~~~~~~~(3.27)

$a_{23}=2\omega(\mu\cosh 2\nu t-\nu\sinh 2\nu t-\mu),$

$a_{31}=-\omega(\mu\cosh 2\nu t+\nu\sinh 2\nu t-\mu),$

$a_{32}=-\omega(\mu\cosh 2\nu t-\nu\sinh 2\nu t-\mu),$

$a_{33}=-2(\omega^2\cosh 2\nu t-\mu^2).$

In the underdamped case $(\mu<\omega), \Omega^2=\omega^2-\mu^2$ we have
$$Te^{Kt}T=-{e^{-2\lambda t}\over 2\Omega^2}\left(\matrix{b_{11}&b_{12}&b_{13}
\cr
b_{21}&b_{22}&b_{23}\cr
b_{31}&b_{32}&b_{33}\cr}\right)$$
with

$b_{11}=(\mu^2-\Omega^2)\cos 2\Omega t-2\mu\Omega\sin 2\Omega t-\omega^2,$

$b_{12}=(\mu^2+\Omega^2)\cos 2\Omega t-\omega^2,$

$b_{13}=2\omega(\mu\cos 2\Omega t-\Omega\sin 2\Omega t-\mu),$

$b_{21}=(\mu^2+\Omega^2)\cos 2\Omega t-\omega^2,$

$b_{22}=(\mu^2-\Omega^2)\cos 2\Omega t+2\mu\Omega\sin 2\Omega t-\omega^2,$
~~~~~~~~~~~~~~~~~(3.28)

$b_{23}=2\omega(\mu\cos 2\Omega t+\Omega\sin 2\Omega t-\mu),$

$b_{31}=-\omega(\mu\cos 2\Omega t-\Omega\sin 2\Omega t-\mu),$

$b_{32}=-\omega(\mu\cos 2\Omega t+\Omega\sin 2\Omega t-\mu),$

$b_{33}=-2(\omega^2\cos 2\Omega t-\mu^2).$

\vskip 1truecm

{\bf 4. The method of the characteristic function}

Instead of solving the master equation (3.7) directly, we first introduce the
normally ordered quantum characteristic function $\chi(\Lambda,\Lambda^*,t)$
defined in terms of the density operator $\rho$ by
$$\chi(\Lambda,\Lambda^*,t)=Tr[\rho(t)\exp(\Lambda a^+)\exp(-\Lambda^*a)],
\eqno(4.1)$$
where $\Lambda$ is a complex variable and the trace is performed over the
states of system. Substituting Eq.(4.1) into the master equation (3.7) and
using the operator relations
$$a\exp(\Lambda a^+)=\exp(\Lambda a^+)(a+\Lambda),$$
$$a^+\exp(-\Lambda^*a)=\exp(-\Lambda^*a)(a^++\Lambda^*),$$
$$Tr[\rho(t)\exp(\Lambda a^+)\exp(-\Lambda^*a)(a^++\Lambda^*)]=\partial_
\Lambda\chi,$$
$$Tr[\rho(t)\exp(\Lambda a^+)\exp(-\Lambda^*a)a]=-\partial_{\Lambda^*}\chi$$
or applying the rules:
$$\rho\leftrightarrow\chi$$
$$a\rho\leftrightarrow -{\partial\over\partial\Lambda^*}\chi$$
$$a^+\rho\leftrightarrow({\partial\over\partial\Lambda}-\Lambda^*)\chi$$
$$\rho a\leftrightarrow(-{\partial\over\partial\Lambda^*}+\Lambda)\chi$$
$$\rho a^+\leftrightarrow{\partial\over\partial\Lambda}\chi,$$
the following partial differential equation for $\chi$ is found:
$$\{\partial_t+[(\lambda-i\omega)\Lambda+\mu\Lambda^*]\partial_\Lambda+
[(\lambda+i\omega)\Lambda^*+\mu\Lambda]\partial_{\Lambda^*}\}\chi(\Lambda,
\Lambda^*,t)=$$
$$=\{L\vert\Lambda\vert^2+C\Lambda^2+C^*\Lambda^{*2}\}\chi(\Lambda,\Lambda^*,t)
,
\eqno(4.2)$$
where
$$L=\lambda-D_2, C={1\over 2}(\mu+D_1^*).$$
We consider the state of the system initially to be a superposition of
coherent states. The coherent states $\vert\alpha>$ of the harmonic
oscillator are minimum uncertainty states having mean coordinate $<q>$ and
mean momentum $<p>$ given by
$$<q>=<\alpha\vert q\vert\alpha>=\sqrt{{2\hbar\over m\omega}}Re\alpha,
 <p>=<\alpha\vert p\vert\alpha>=\sqrt{2\hbar m\omega}Im\alpha.\eqno(4.3)$$
Consequently, we take as the initial density operator
$$\rho(0)=\int d\alpha d\beta N(\alpha,\beta)\vert\alpha><\beta\vert.$$
The quantum characteristic function corresponding to the operator
$\vert\alpha><\beta\vert$ is given from Eq. (4.1) by
$$\chi=<\beta\vert\alpha>\exp(\lambda\beta^*-\lambda^*\alpha).\eqno(4.4)$$
We look for a solution of (4.2) having the exponential form
$$\chi(\Lambda,\Lambda^*,t)=
\int d\alpha d\beta N(\alpha,\beta)<\beta\vert\alpha>\exp[A(t)\Lambda+B(t)
\Lambda^*+f(t)\Lambda^2+f^*(t)\Lambda^{*2}+h(t)\vert\Lambda\vert^2].\eqno(4.5)
$$
The form of the solution (4.5) is suggested from the fact that the left-hand
side of Eq. (4.2) contains first-order derivatives with respect to the time
and variables $\Lambda$ and $\Lambda^*$ and is symmetric with
respect to complex conjugation.
The functions $A(t),B(t),f(t)$ and $h(t)$ depend only on time. Corresponding
to the initial factor in Eq. (4.4), these functions have to satisfy the
initial
conditions
$$A(0)=\beta^*, B(0)=\alpha, f(0)=0,h(0)=0.\eqno(4.6)$$
When we introduce the function (4.5) into Eq. (4.2) and equate the
coefficients
for equal powers of $\Lambda$ and $\Lambda^*$, we get the following two
systems of linear differential equations of first order with constant
coefficients:
$${dA(t)\over dt}+(\lambda-i\omega)A(t)+\mu B(t)=0$$
$${dB(t)\over dt}+\mu A(t)+(\lambda+i\omega)B(t)=0\eqno(4.7a)$$
$${dR(t)\over dt}+2\lambda R(t)+2\omega I(t)+\mu h(t)=Re C$$
$${dI(t)\over dt}+2\lambda I(t)-2\omega R(t)=Im C\eqno(4.7b)$$
$${dh(t)\over dt}+4\mu R(t)+2\lambda h(t)=L,$$
where $R(t)=Re f(t),I(t)=Im f(t)$ with the initial conditions $R(0)=I(0)=
h(0)=0$.
Subject to the initial conditions (4.6), the homogeneous system $(4.7a)$ has
the
solution [19]:
$$A(t)=u(t)\beta^*-v(t)\alpha,$$
$$B(t)=-u^*(t)\alpha+v(t)\beta^*,\eqno(4.8)$$
where
$$u(t)={1\over 2}[\exp(-\mu_-t)+\exp(-\mu_+t)+{2i\omega\over\mu_--\mu_+}
(\exp(-\mu_+t)-\exp(-\mu_-t))],$$
$$v(t)={\mu\over\mu_--\mu_+}(\exp(-\mu_-t)-\exp(-\mu_+t)).\eqno(4.9)$$
The eigenvalues $\mu_{\pm}$ are given by
$$\mu_{\pm}=\lambda\pm\sqrt{\mu^2-\omega^2}, \gamma\equiv\sqrt{\mu^2-\omega^2}.
\eqno(4.10)$$
The system $(4.7b)$ has the eigenvalues $-2\lambda, -2(\lambda\pm\sqrt{\mu^2-
\omega^2})=-2\mu_{\pm}$ and in order to integrate it we apply the
same method as for the system (3.13) in the preceding Section. We obtain:
$$f(t)={P\over 2\mu}\exp(-2\mu_+t)(\gamma-i\omega)-{N\over 2\mu}\exp(-2\mu_-t)
(\gamma+i\omega)-{i\mu M\over 2\omega}\exp(-2\lambda t)+f(\infty),$$
$$h(t)=M\exp(-2\lambda t)+N\exp(-2\mu_-t)+P\exp(-2\mu_+t)+h(\infty).\eqno(4.11)
$$
Here $M,N,P,f(\infty)$ and $h(\infty)$ are constants given by [19]:
$$M={\omega\over\lambda\gamma^2}(\mu Im C+{1\over 2}\omega L),$$
$$N={\mu\over 2\gamma^2(\lambda-\gamma)}(\gamma Re C-\omega Im C-{\mu L\over 2}
),$$
$$P=-{\mu\over 2\gamma^2(\lambda+\gamma)}(\gamma Re C+\omega Im C+{\mu L\over 2
})$$
and the asymptotic values connected with the diffusion coefficients $D_{qq},
D_{pp}$ and $D_{pq}$ are:
$$R(\infty)={2(\lambda Re C-\omega Im C)-L\mu\over 4(\lambda^2-\gamma^2)},$$
$$I(\infty)={2\omega\lambda Re C+2(\lambda^2-\mu^2)Im C-L\mu\omega\over 4
\lambda(\lambda^2-\gamma^2)},\eqno(4.12)$$
$$h(\infty)={L(\lambda^2+\omega^2)-2\mu(\lambda Re C-\omega Im C)\over 2
\lambda(\lambda^2-\gamma^2)}.$$

By knowing the characteristic function (4.5), (4.8)-(4.11) corresponding to
the initial density operator which represents a superposition of coherent
states, it is easy to obtain explicit formulae for the moments:
$$<a^{+m}(t)a^n(t)>=Tr[a^{+m}(t)a^n(t)\rho(t)]=(-1)^n{\partial^{n+m}\over
\partial\Lambda^{*n}\partial\Lambda^m}\chi(\Lambda,\Lambda^*,t)\vert_{
\Lambda=\Lambda^*=0}.$$

In the following, we take the density operator $\rho$ in the coherent state
representation
$$\rho(0)=\int P(\alpha)\vert\alpha><\alpha\vert d^2\alpha,$$
where $P(\alpha)$ is the diagonal or $P$ Glauber distribution and $d^2\alpha=
dRe\alpha dIm\alpha$. The integration covers the entire complex $\alpha$
plane. Then the characteristic function (4.5) becomes:
$$\chi(\Lambda,\Lambda^*,t)=\int d^2\alpha P(\alpha)\exp[(u\alpha^*-v\alpha)
\Lambda+(-u^*\alpha+v\alpha^*)\Lambda^*]\exp[f\Lambda^2+f^*\Lambda^{*2}+
h\vert\Lambda\vert^2].$$
Let us assume that the damped oscillator is at $t=0$ prepared in a pure
coherent state, say $\vert\alpha_0>$, corresponding to $P(\alpha)=\delta(Re
\alpha-Re\alpha_0)\delta(Im\alpha-Im\alpha_0)$. One has
$$\chi^{(0)}(\Lambda,\Lambda^*,t)=\exp[(u\alpha_0^*-v\alpha_0)\Lambda+(-u^*
\alpha_0+v\alpha_0^*)\Lambda^*]\exp[f\Lambda^2+f^*\Lambda^{*2}+h\vert\Lambda
\vert^2].$$

The first moments are given by
$$<a^+(t)>={\partial\chi^{(0)}(t)\over\partial\Lambda}\vert_{\Lambda=\Lambda^*
=0}=u\alpha_0^*-v\alpha_0,$$
$$<a(t)>=-{\partial\chi^{(0)}(t)\over\partial\Lambda^*}\vert_{\Lambda=\Lambda^*
=0}=u^*\alpha_0-v\alpha_0^*.$$
Then, with the notations (3.9),using (4.3) and the transformations
$$q(t)=\sqrt{{\hbar\over 2m\omega}}(a^+(t)+a(t)),$$
$$p(t)=i\sqrt{{\hbar m\omega\over 2}}(a^+(t)-a(t))$$
for the displacement operator $q(t)$ and the momentum operator $p(t)$ of the
oscillator, we obtain the following mean values:
$$\sigma_q(t)=\sqrt{{\hbar\over 2m\omega}}((u-v)\alpha_0^*+(u^*-v)\alpha_0),$$
$$\sigma_p(t)=i\sqrt{{\hbar m\omega\over 2}}((u+v)\alpha_0^*-(u^*+v)\alpha_0),
$$
with $u,v$ given by (4.9), (4.10).
There are two cases:

$a)$ the overdamped case: $\mu>\omega,\nu^2=\mu^2-\omega^2,\gamma\equiv\nu$;
then
$$u(t)=\exp(-\lambda t)(\cosh\nu t+{i\omega\over\nu}\sinh\nu t),$$
$$v(t)=-{\mu\over\nu}\exp(-\lambda t)\sinh\nu t\eqno(4.13)$$
and $\sigma_q(t),\sigma_p(t)$ take the previous form (3.15);

$b)$ the underdamped case: $\mu<\omega,\Omega^2=\omega^2-\mu^2,\gamma\equiv
i\Omega$;
then
$$u(t)=\exp(-\lambda t)(\cosh\Omega t+{i\omega\over\Omega}\sinh\Omega t),$$
$$v(t)=-{\mu\over\Omega}\exp(-\lambda t)\sinh\Omega t\eqno(4.14)$$
and $\sigma_q(t),\sigma_p(t)$ take the previous form (3.16).

For the variances one finds:
$$<a^2(t)>={\partial^2\chi^{(0)}(t)\over\partial\Lambda^{*2}}\vert_{\Lambda=
\Lambda^*
=0}=(u^*\alpha_0-v\alpha_0^*)^2+2f^*,$$
$$<a^{+2}(t)>={\partial^2\chi^{(0)}(t)\over\partial\Lambda^2}\vert_{\Lambda=
\Lambda^*
=0}=(u\alpha_0^*-v\alpha_0)^2+2f,\eqno(4.15)$$
$$<a^+(t)a(t)>=-{\partial^2\chi^{(0)}(t)\over\partial\Lambda\partial\Lambda^*}
\vert_
{\Lambda=\Lambda^*=0}=(u\alpha_0^*-v\alpha_0)(u^*\alpha_0-v\alpha_0^*)-h.$$
Then the relations (3.9) will give us the explicit time dependence of the
variances $\sigma_{qq}(t)$,
 $\sigma_{pp}(t)$, $\sigma_{pq}(t)$. The asymptotic
values of these variances are given by the following expresions [19]:
$$\sigma_{qq}(\infty)={\hbar\over m\omega}(f+f^*-h+{1\over 2})\vert_{t\to
\infty},$$
$$\sigma_{pp}(\infty)=-\hbar m\omega(f+f^*+h-{1\over 2})\vert_{t\to\infty},$$
$$\sigma_{pq}(\infty)=i\hbar(f-f^*)\vert_{t\to\infty}.$$
With $f(\infty)=R(\infty)+iI(\infty)$ and using the formulas (4.12) for
$R(\infty),I(\infty),h(\infty)$, the asymptotic values of the variances take
the same form (3.20) as in the preceding Section, as expected.

With the relations (4.15), the expectation value of the energy operator can be
calculated:
$$E(t)=\hbar\omega(<a^+a>+{1\over 2})+{i\hbar\mu\over 2}(<a^{+2}>-<a^2>).$$
The asymptotic mean value of the energy of the open harmonic oscillator is:
$$E(\infty)={1\over 2m}\sigma_{pp}(\infty)+{1\over 2}m\omega^2\sigma_{qq}
(\infty)+\mu\sigma_{pq}(\infty),$$
or, as a function of diffusion coefficients, by using (3.20):
$$E(\infty)={1\over\lambda}({1\over 2m}D_{pp}+{m\omega^2\over 2}D_{qq}+\mu D_
{pq}).$$

\vskip 1truecm

{\bf 5. Quasiprobability distributions for damped harmonic oscillator}

The methods of quasiprobabilities have provided technical tools of great power
for the statistical description  of microscopic systems formulated in terms of
the density operator [40-42,58]. The first quasiprobability method was that
introduced by Wigner [43] in a quantum-mechanical context. In quantum optics
the $P$ representation introduced by Glauber [44,45] and Sudarshan [46]
provided many practical applications of quasiprobabilities. The development
of quantum-mechanical master equations was combined with the Glauber $P$
representation to give a Fokker-Planck equation for the laser [47,48]. One
useful way to study the consequences of the master equation for the
one-dimensional damped harmonic oscillator is to transform it into equations
for the $c$-number quasiprobability distributions associated with the density
operator. The resulting differential equations of the Fokker-Planck type for
the distribution functions can be solved by standard methods and observables
directly calculated as correlations of these distribution functions. However,
the Fokker-Planck equations do not always have positive-definite diffusion
coefficients. In this case one can treat the problem with the generalized
$P$ distribution [41].

First we present a short summary of the theory of quasiprobability
distributions. For the master equation (3.7) of the harmonic oscillator,
physical observables can be obtained from the expectation values of
polynomials of the annihilation and creation operators. The expectation values
are determined by using the quantum density operator $\rho$. Usually one
expands the density operator with the aid of coherent states, defined as
eigenstates of the annihilation operator: $a\vert\alpha>=\alpha\vert\alpha>$.
They are given in terms of the eigenstates of the harmonic oscillator as
$$\vert\alpha>=\exp(-{\vert\alpha\vert^2\over 2})\sum_{n=0}^\infty{1\over
\sqrt{n!}}\alpha^n\vert n>,$$
with the normalization $\vert<\beta\vert\alpha>\vert^2=\exp(-\vert\alpha-\beta
\vert^2)$. In order to solve the master equation (3.7) we represent the density
operator $\rho$ by a distribution function over a $c$-number phase space.
The chosen distribution function, introduced in [49] is defined as follows:
$$\Phi(\alpha,s)={1\over\pi^2}\int\chi(\Lambda,s)\exp(\alpha\Lambda^*-\alpha
^*\Lambda)d^2\Lambda, \eqno(5.1)$$
with the characteristic function
$$\chi(\Lambda,s)=Tr[\rho D(\Lambda,s)],$$
where $D(\Lambda,s)$ is the displacement operator
$$D(\Lambda,s)=\exp(\Lambda a^+-\Lambda^*a+{1\over 2}s\vert\Lambda\vert^2).$$
The interval of integration in Eq. (5.1) is the whole complex $\Lambda$ plane.
Because of
$$\delta^2(\alpha)={1\over\pi^2}\int\exp(\alpha\Lambda^*-\alpha^*\Lambda)d^2
\Lambda,$$
the characteristic function $\Phi(\alpha,s)$ is the Fourier transform of the
characteristic function. Since the density operator is normalized by $Tr\rho=
1$, one obtains the normalization of $\Phi$:
$$\int\Phi(\alpha,s)d^2\alpha=1.$$
In this paper we restrict ourselves to distribution functions with the
parameters $s=1,0$ and $-1$. These distribution functions can be used to
calculate expectation values of products of annihilation and creation
operators. For that purpose one first expands the displacements operator in a
power series of the operators $a$ and $a^+$:
$$D(\Lambda,s)=\sum_{m=0}^\infty\sum_{n=0}^\infty{\Lambda^m(-\Lambda^*)^n\over
 m!n!}\{a^{+m}a^n\}_s.\eqno(5.2)$$
The braces with the index $s$ indicate the special representations of the
polynomials depending on $s$. For example for $n=m=1$ we have the $s$-ordered
operators:
$$\{a^+a\}_{s=1}=a^+a,$$
$$\{a^+a\}_{s=0}={1\over 2}(a^+a+aa^+),\eqno(5.3)$$
$$\{a^+a\}_{s=-1}=aa^+.$$
Expectation values of the $s$-ordered operators can be calculated as follows:
$$<\{a^{+m}a^n\}_s>=Tr[\rho\{a^{+m}a^n\}_s]=Tr[\rho({\partial\over\partial
\Lambda})^m(-{\partial\over\partial\Lambda^*})^nD(\Lambda,s)\vert_{\Lambda=0}]=
$$
$$=({\partial\over\partial\Lambda})^m(-{\partial\over\partial\Lambda^*})^n\chi
(\Lambda,s)\vert_{\Lambda=0}=\int(\alpha^*)^m\alpha^n\Phi(\alpha,s)d^2\alpha.$$
For the last step we apply the inverse relation to Eq. (5.1):
$$\chi(\Lambda,s)=\int\Phi(\alpha,s)\exp(\Lambda\alpha^*-\Lambda^*\alpha)d^2
\alpha.$$
In the following we discuss the distribution functions for $s=1,0$ and $-1$ in
more detail. For $s=1$ we obtain the Glauber $P$ function [44,45,50], for $s=0$
the Wigner function [43] and for $s=-1$ the $Q$ function [49]. For $s=1$ we
have
$$D(\Lambda,1)=\exp(\Lambda a^+)\exp(-\Lambda^*a).$$
Then the $s$ ordering in Eq. (5.2) corresponds to normal ordering. Since the
Glauber $P$ function is the Fourier transform of the characteristic function
$$\chi_N(\Lambda)=Tr[\rho\exp(\Lambda a^+)\exp(-\Lambda^*a)]=\chi(\Lambda,1),$$
it follows from Eq. (5.1) that the distribution $\Phi(\alpha,1)$ is identical
to the $P$ function. This function is used for an expansion of the density
operator in diagonal coherent state projection operators [44-46,51]:
$$\rho=\int P(\alpha)d^2\alpha\vert\alpha><\alpha\vert.$$
Calculating the expectation value of normally ordered operator products we
obtain the relation
$$<a^{+m}a^n>=\int(\alpha^*)^m\alpha^n\Phi(\alpha,1)d^2
\alpha=\int(\alpha^*)^m\alpha^nP(\alpha)d^2\alpha,$$
from which we again derive $P(\alpha)=\Phi(\alpha,1)$. Despite the formal
similarity to averaging with a classical probability distribution, the function
$P(\alpha)$ is not a true probability distribution. Because of the
overcompleteness of the coherent states, the $P$ function is not a unique,
well-behaved positive function for all density operators.

Cahill [52] studied the $P$ representation for density operators which
represent pure states and found a narrow class of states for which the $P$
representation exists. They can be generated from a particular coherent state
$\vert\alpha>$ by the application of a finite number of creation operators.
For example, for the ground state of the harmonic oscillator it is easy to
show that $\chi_N(\Lambda)=1$ for all $\Lambda$. In that case the $P$ function
becomes $P(\alpha)=\delta^2(\alpha)$. The delta function and its derivatives
are examples of a class of generalized functions known as tempered
distributions [50]. Also Cahill [53] introduced a representation of the
density operator of the electromagnetic field that is suitable for all density
operators and that reduces to the coherent state $P$ representation when the
latter exists. The representation has no singularities.

Sudarshan [46] offered a singular formula for the $P$ representation in terms
of an infinite series of derivatives of the delta function. From the
mathematical point of view, such a series is usually not considered to be a
distribution function [50,51].

For $s=-1$ we have
$$D(\Lambda,-1)=\exp(-\Lambda^*a)\exp(\Lambda a^+).$$
The $s$ ordering corresponds to antinormal ordering. Because the $Q$ function
is the Fourier transform of the characteristic function
$$\chi_A(\Lambda)=Tr[\rho\exp(-\Lambda^*a)\exp(\Lambda a^+)]=\chi
(\Lambda,-1),$$
it follows from Eq. (5.1) that the distribution $\Phi(\alpha,-1)$ is the $Q$
function. It is given by the diagonal matrix elements of the density operator
in terms of coherent states:
$$Q(\alpha)={1\over\pi}<\alpha\vert\rho\vert\alpha>.$$
Though for all density operators the $Q$ function is bounded, non-negative and
infinitely differentiable, it has the disadvantage that not every positive $Q$
function corresponds to a positive semidefinite Hermitian density operator.
Evaluating moments is only simple in the $Q$ representation for antinormally
ordered operator products.

For $s=0$, the distribution $\Phi(\Lambda,0)$ becomes the Wigner function $W$.
The latter function is defined as the Fourier transform of the characteristic
function
$$\chi_S(\Lambda)=Tr[\rho\exp(\Lambda a^+-\Lambda^*a)]=\chi(\Lambda,0).$$
Because this characteristic function is identical to $\chi(\Lambda,0)$, we
conclude that $\Phi(\alpha,0)$ is the Wigner function $W(\alpha)$. Therefore,
the Wigner function can be used to calculate expectation values of
symmetrically ordered operators:
$$<\{a^{+m}a^n\}_{s=0}>=\int(\alpha^*)^m\alpha^nW(\alpha)d^2\alpha.$$
The symmetrically ordered operators are the arithmetic average of $(m+n)!
/(m!n!)$ differently ordered products of $m$
factors of $a^+$ and $n$ factors of $a$. An example for $m=n=1$ is given in
Eq. (5.3).

The Wigner function is a nonsingular, uniformly continuous function of
$\alpha$ for all density operators and may in general assume negative values.
It is related to the density operator as follows:
$$W(\alpha)={1\over\pi^2}\int d^2\Lambda Tr[\exp(\Lambda(a^+-\alpha^*)-\Lambda
^*(a-\alpha))\rho].$$
Also it can be obtained from the $P$ representation:
$$W(\alpha)={2\over\pi}\int P(\beta)\exp(-2\vert\alpha-\beta\vert^2)d^2\beta.$$

By using the standard transformations [54,55], the master equation (3.7) can be
transformed into a differential equation for a corresponding $c$-number
distribution. In this Section we apply these methods to derive Fokker-Planck
equations for the before mentioned distributions: the Glauber $P$, the $Q$ and
Wigner distributions. Using the relations
$${\partial D(\Lambda,s)\over\partial\Lambda}=[(s-1){\Lambda^*\over 2}+a^+]
D(\Lambda,s)=D(\Lambda,s)[(s+1){\Lambda^*\over 2}+a^+],$$
$${\partial D(\Lambda,s)\over\partial\Lambda^*}=[(s+1){\Lambda\over 2}-a]
D(\Lambda,s)=D(\Lambda,s)[(s-1){\Lambda\over 2}-a],$$
one can derive the following rules for transforming the master equation (3.7)
into Fokker-Planck equations in the Glauber $P(s=1)$, the $Q(s=-1)$
and Wigner $W(s=0)$ representations:
$$a\rho\leftrightarrow (\alpha-{s-1\over 2}{\partial\over\partial\alpha^*})
\Phi$$
$$a^+\rho\leftrightarrow (\alpha^*-{s+1\over 2}{\partial\over\partial\alpha})
\Phi$$
$$\rho a\leftrightarrow (\alpha-{s+1\over 2}{\partial\over\partial\alpha^*})
\Phi$$
$$\rho a^+\leftrightarrow (\alpha^*-{s-1\over 2}{\partial\over\partial\alpha})
\Phi.$$

Applying these operator correspondences (repeatedly, if necessary), we
find the following Fokker-Planck equations for the distributions
$\Phi(\alpha,s)$:
$${\partial\Phi(\alpha,s)\over\partial t}=-({\partial\over\partial\alpha}d_
\alpha+{\partial\over\partial\alpha^*}d_\alpha^*)\Phi(\alpha,s)+$$
$$+{1\over 2}({\partial^2\over\partial\alpha^2}D_{\alpha\alpha}+{\partial^2
\over\partial \alpha^{*2}}D_{\alpha\alpha}^*+2{\partial^2\over\partial\alpha
\partial\alpha^*}D_{\alpha\alpha^*})\Phi(\alpha,s).\eqno(5.4)$$
Here, $\Phi(\alpha,s)$ is $P(s=1), Q(s=-1)$ or $W(s=0)$. While the drift
coefficients are the same for the three distributions, the diffusion
coefficients are different:
$$d_\alpha=-(\lambda+i\omega)\alpha+\mu\alpha^*,
 D_{\alpha\alpha}=D_1+s\mu,
 D_{\alpha\alpha^*}=D_2-s\lambda.$$

The Fokker-Planck equation (5.4) can also be written in terms of real
coordinates $x_1$ and $x_2$ defined by
$$\alpha=x_1+ix_2\equiv\sqrt{{m\omega\over 2\hbar}}<q>+i{1\over\sqrt{2\hbar
 m\omega}}<p>,$$
$$\alpha^*=x_1-ix_2\eqno(5.5)$$
as follows:
$${\partial\Phi(x_1,x_2)\over\partial t}=-({\partial\over\partial x_1}d_1+
{\partial\over\partial x_2}d_2)\Phi(x_1,x_2)+$$
$$+{1\over 2}({\partial^2\over\partial x_1^2}D_{11}+{\partial^2\over
\partial x_2^2}D_{22}+2{\partial^2\over
\partial x_1\partial x_2}D_{12})\Phi(x_1,x_2),\eqno(5.6)$$
with the new drift and difusion coefficients given by
$$d_1=-(\lambda-\mu)x_1+\omega x_2,
d_2=-\omega x_1-(\lambda+\mu)x_2,$$
$$D_{11}={1\over\hbar}m\omega D_{qq}-{s\over 2}(\lambda-\mu),
 D_{22}={1\over\hbar}{D_{pp}\over m\omega}-{s\over 2}(\lambda+\mu),
 D_{12}={1\over\hbar}D_{pq}.$$
We note that the diffusion matrix
$$D=\left(\matrix{D_{11}&D_{12}\cr
D_{12}&D_{22}\cr}\right)$$
for the $P$ ddistribution ($s=1$) needs not to be positive definite.

Since the drift coefficients are linear in the variables $x_1$ and $x_2
 (i=1,2)$:
$$d_i=-\sum_{j=1}^2 A_{ij}x_j, A_{ij}=-{\partial d_i\over\partial x_j},$$
with
$$A=\left(\matrix{\lambda-\mu&-\omega\cr
\omega&\lambda+\mu\cr}\right)\eqno(5.7)$$
and the diffusion coefficients are constant with respect to $x_1$ and $x_2$,
Eq. (5.6) describes an Ornstein-Uhlenbeck process [56,57].

The solution of the Fokker-Planck equation (5.6) can immediately be written
down provided that the diffusion matrix $D$ is positive definite. However,
the diffusion matrix in the Glauber $P$ representation is not, in general,
positive definite. For example, if
$$D_{11}^PD_{22}^P-(D_{12})^2<0,$$
the $P$ distribution does not exist as a well-behaved function. In this
situation, the so-called generalized $P$ distributions can be taken that are
well-behaved, normal ordering functions [41]. The $Q$ and $W$ distributions
always exist; they are Gaussian functions if they are initially of Gaussian
type.

From Eq. (5.6) one can directly derive the equations of motion for the
expectation values of the variables $x_1$ and $x_2 (i=1,2)$:
$${d<x_i>\over dt}=-\sum_{j=1}^2 A_{ij}<x_j>.\eqno(5.8)$$
By using Eqs. (3.5), (5.5) and (5.8)
we obtain the equations of motion for the
expectation values $\sigma_q(t),\sigma_p(t)$  of coordinate and momentum of the
harmonic oscillator which are identical with those derived in the preceding
two Sections by using the Heisenberg representation and the method of
characteristic function, respectively (see Eqs. (3.12)).

The variances of the variables $x_1$ and $x_2$ are defined by the expectation
values
$$\sigma_{ij}=<x_ix_j>-<x_i><x_j>, ij=1,2.$$
They are connected with the variances and covariance of the coordinate $q$ and
momentum $p$ by
$$\sigma_{qq}=(2\hbar/m\omega)\sigma_{11}, \sigma_{pp}=2\hbar m\omega\sigma_{22
},$$
$$\sigma_{pq}=<{1\over 2}(pq+qp)>-<p><q>=2\hbar\sigma_{12}.$$
They can be calculated with the help of the variances of the quasiprobability
distributions $(i,j=1,2)$:
$$\sigma_{ij}^{(s)}=\int x_ix_j\Phi(x_1,x_2,s)dx_1dx_2-\int x_i\Phi(x_1,x_2,s)
dx_1dx_2\int x_j\Phi(x_1,x_2,s)dx_1dx_2.$$
The following relations exist between the various variances:
$$\sigma_{ii}=\sigma_{ii}^P+{1\over 4}=\sigma_{ii}^Q-{1\over 4}=\sigma_{ii}^W,
 i=1,2,$$
$$\sigma_{12}=\sigma_{12}^P=\sigma_{12}^Q=\sigma_{12}^W.$$
The variances $\sigma_{ij}^{(s)}$ fulfill the following equations of motion:
$${d\sigma_{ij}^{(s)}\over dt}=-\sum_{l=1}^2(A_{il}\sigma_{lj}^{(s)}+\sigma
_{il}^{(s)}A_{lj}^T)+D_{ij}^{(s)}.\eqno(5.9)$$
They can be written explicitly in the form:
$${d\sigma_{11}^{(s)}\over dt}=-2A_{11}\sigma_{11}^{(s)}-2A_{12}\sigma_{12}
^{(s)}+D_{11}^{(s)},$$
$${d\sigma_{22}^{(s)}\over dt}=-2A_{21}\sigma_{12}^{(s)}-2A_{22}\sigma_{22}
^{(s)}+D_{22}^{(s)},$$
$${d\sigma_{12}^{(s)}\over dt}=-(A_{11}+A_{22})\sigma_{12}^{(s)}-A_{21}\sigma
_{11}^{(s)}-A_{12}\sigma_{22}^{(s)}+D_{12}^{(s)},$$
where the matrix elements $A_{ij}$ are defined in Eq. (5.7). These relations
are
sufficient to prove  that the equations of motion for the variances
$\sigma_{11}$ and $\sigma_{22}$ and the covariance $\sigma_{12}$ are the same
irrespective of the choice of the representation as expected. The
corresponding
equations of motion of the  variances and covariance of the coordinate and
momentum coincide with those obtained in the preceding two Sections by using
the Heisenberg representation and the method of characteristic function,
respectively (see Eqs. (3.13)).

In order that the system approaches a steady state, the condition $\lambda>\nu$
must be met. Thus the steady-state solutions are
$$\Phi(x_1,x_2,s)={1\over 2\pi\sqrt{det(\sigma(\infty))}}\exp[-{1\over 2}
\sum_{i,j=1,2}(\sigma^{-1})_{ij}(\infty)x_ix_j],\eqno(5.10)$$
where the stationary covariance matrix
$$\sigma(\infty)=\sigma^{(s)}(\infty)=\left(\matrix{\sigma_{11}^{(s)}(\infty)&
\sigma_{12}^{(s)}(\infty)\cr
\sigma_{12}^{(s)}(\infty)&\sigma_{22}^{(s)}(\infty)\cr}\right)$$
can be determined from the algebraic equation (see Eq. (5.9)):
$$\sum_{l=1}^2(A_{il}\sigma_{lj}^{(s)}(\infty)+\sigma_{il}^{(s)}(\infty)A_{lj}
^T)=D_{ij}^{(s)}.$$
With the matrix elements $A_{ij}$ given by (5.7), we obtain
$$\sigma_{11}^{(s)}(\infty)={(2\lambda(\lambda+\mu)+\omega^2)D_{11}^{(s)}+
\omega^2D_{22}^{(s)}+2\omega(\lambda+\mu)D_{12}^{(s)}\over 4\lambda(\lambda^2+
\omega^2-\mu^2)},$$
$$\sigma_{22}^{(s)}(\infty)={\omega^2D_{11}^{(s)}+(2\lambda(\lambda-\mu)+
\omega^2)D_{22}^{(s)}-2\omega(\lambda-\mu)D_{12}^{(s)}\over 4\lambda(\lambda^2+
\omega^2-\mu^2)},$$
$$\sigma_{12}^{(s)}(\infty)={-\omega(\lambda+\mu)D_{11}^{(s)}+\omega(\lambda-
\mu)D_{22}^{(s)}+2(\lambda^2-\mu^2)D_{12}^{(s)}\over 4\lambda(\lambda^2+\omega
^2-\mu^2)}.$$
The explicit matrix elements $\sigma_{ij}^{(s)}$ for the three representations
$P,Q$ and $W$ can be obtained by inserting the corresponding diffusion
coefficients.
The distribution functions (5.10) can be used to calculate the
expectation values of the coordinate and momentum and the variances by direct
integration. The following relations are noticed [37]:
$$\sigma_{ij}^W(\infty)={1\over 2}(\sigma_{ij}^P(\infty)+\sigma_{ij}^Q(\infty)
), i,j=1,2.$$
The uncertainty principle $\sigma_{11}\sigma_{22}\ge 1/16$ gives rise to
the conditions $4\sigma_{11}^Q\sigma_{22}^Q\ge\sigma_{11}^Q+\sigma_{22}^Q$
and $\sigma_{11}^W\sigma_{22}^W\ge 1/16$ for the $Q$ and $W$
distributions, respectively.

\vskip 1truecm

{\bf 6. Density matrix of the damped harmonic oscillator}

In this section we explore the general results that follow from the master
equation of the one-dimensional damped harmonic oscillator. Namely, we discuss
the physically relevant solutions of the master equation, by using the method
of the generating function. In particular, we provide extended solutions
(including both diagonal and off-diagonal matrix elements) for different
initial conditions.

The method used in this section follows closely  the procedure of Jang [39].
Let us first rewrite the master equation (3.7) for the density matrix by means
of the number representation. Specificallly, we take the matrix elements of
each term between different number states denoted by $\vert n>$, and using
$a^+\vert n>=\sqrt{n+1}\vert n+1>$ and $a\vert n>=\sqrt n\vert n-1>$ we get
$${d\rho_{mn}\over dt}=-i\omega(m-n)\rho_{mn}+\lambda\rho_{mn}-(m+n+1)D_2\rho
_{mn}+$$
$$+{1\over 2}\sqrt{m(m-1)}(D_1+\mu)\rho_{m-2,n}
-\sqrt{m(n+1)}D_1\rho_{m-1,n+1}+$$
$$+{1\over 2}\sqrt{(n+1)(n+2)}(D_1-\mu)\rho
_{m,n+2}
+{1\over 2}\sqrt{(m+1)(m+2)}(D_1^*-\mu)\rho_{m+2,n}-$$
$$-\sqrt{m+1)n}D_1^*\rho
_{m+1,n-1}
+{1\over 2}\sqrt{(n-1)n}(D_1^*+\mu)\rho_{m,n-2}+$$
$$+\sqrt{(m+1)(n+1)}(D_2+
\lambda)\rho_{m+1,n+1}+\sqrt{mn}(D_2-\lambda)\rho_{m-1,n-1}.\eqno(6.1)$$
Here, we have used the abbreviated notation
$$\rho_{mn}=<m\vert\rho(t)\vert n>.$$
This master equation is complicated in form and in indices involved. It
comprises not only the density matrix in symmetrical forms, such as
$\rho_{m\pm1,n\pm1}$, but also those matrix elements in asymmetrical forms like
$\rho_{m\pm2,n}, \rho_{m,n\pm2}$ and $\rho_{m\mp1,n\pm1}$.
In order to solve Eq. (6.1) we use the method of a generating function which
allows us to
eliminate the variety of indices $m$ and $n$ implicated in the equation. When
we define the double-fold generating function by
$$G(x,y,t)=\sum_{m,n}{1\over\sqrt{m!n!}}x^my^n\rho_{mn}(t),\eqno(6.2)$$
the density matrix can be evaluated from the inverse relation of Eq. (6.2):
$$\rho_{mn}(t)={1\over\sqrt{m!n!}}({\partial\over\partial x})^m({\partial
\over\partial y})^nG(x,y,t)\vert_{x=y=0},\eqno(6.3)$$
provided that the generating function is calculated beforehand. When we
multiply both sides of Eq. (6.1) by $x^my^n/\sqrt{m!n!}$ and sum over the
result, we get a linear second order partial differential equation for
$G(x,y,t)$, namely
$${\partial\over\partial t}G(x,y,t)=\{[-(i\omega+D_2)x-D_1^*y]{\partial\over
\partial x}+[-D_1x+(i\omega-D_2)y]{\partial\over\partial y}+$$
$$+(D_2+\lambda){\partial^2\over\partial x\partial y}+{1\over 2}[(D_1^*-\mu)
{\partial^2\over\partial x^2}+(D_1-\mu){\partial^2\over\partial y^2}]+$$
$$+[{1\over 2}(D_1+\mu)x^2+{1\over 2}(D_1^*+\mu)y^2+(D_2-\lambda)(xy-1)]\}
G(x,y,t).\eqno(6.4)$$
A special solution of Eq. (6.4) can be taken as
$$G(x,y,t)={1\over A}\exp\{xy-[B(x-C)^2+D(y-E)^2+F(x-C)(y-E)]/H\},\eqno(6.5)$$
where $A,B,C,D,E,F$ and $H$ are unknown functions of time which are to be
determined. When we first substitute the expression (6.5) for $G(x,y,t)$ into
Eq. (6.4) and  equate the coefficients of equal powers of $x,y$ and $xy$ on
both sides of the equation, we get the following differential equations
for the functions $A,B,D,F$ and $H$:
$$-{1\over A}{dA\over dt}=-(D_1^*-\mu){B\over H}-(D_1-\mu){D\over H}-(D_2+
\lambda){F\over H}+2\lambda,\eqno(6.6)$$
$${d\over dt}({B\over H})=2(\lambda-i\omega){B\over H}-\mu{F\over H}-{1
\over 2}(D_1-\mu){F^2\over H^2}-2(D_2+\lambda){FB\over H^2}-2(D_1^*-\mu){B^2
\over H^2},\eqno(6.7)$$
$${d\over dt}({D\over H})=2(\lambda+i\omega){D\over H}-\mu{F\over H}-{1\over 2}
(D_1^*-\mu){F^2\over H^2}-2(D_2+\lambda){DF\over H^2}-2(D_1-\mu){D^2\over H^2},
\eqno(6.8)$$
$${d\over dt}({F\over H})=2\lambda{F\over H}-(D_2+\lambda){F^2\over H^2}-
2\mu({B\over H}+{D\over H})-4(D_2+\lambda){DB\over H^2}-2(D_1^*-\mu){BF\over
H^2}-2(D_1-\mu){DF\over H^2}.\eqno(6.9)$$
In addition to these equations, we get for the functions $C$ and $E$
$$2B{dC\over dt}+F{dE\over dt}=(-2(\lambda-i\omega)B+\mu F)C+(2\mu B-(\lambda+i
\omega)F)E,\eqno(6.10)$$
$$2D{dE\over dt}+F{dC\over dt}=(-2(\lambda+i\omega)D+\mu F)E+(2\mu D-(\lambda-
i\omega)F)C.\eqno(6.11)$$
The equations (6.10) and (6.11) can be reformulated in order to eliminate the
functions $B,D$ and $F$, provided $BD-F^2/4\not=0$. We obtain
$${dC\over dt}=-(\lambda-i\omega)C+\mu E,\eqno(6.12)$$
$${dE\over dt}=-(\lambda+i\omega)E+\mu C.\eqno(6.13)$$
The functions $A,B,D,F$ and $H$ are connected by the auxiliary condition that
$Tr\rho$ is independent of time. The trace of $\rho$ can be evaluated by
summing the diagonal matrix elements $\rho_{nn}$ given in Eq. (6.3) or directly
by using the integral expression
$$Tr\rho=\sum_{n=0}^\infty \rho_{nn}={1\over(2\pi)^2}\int\exp(-k_1k_2)\exp
(ik_1x+ik_2y)G(x,y,t)dk_1dk_2dxdy.$$
We obtain with the generating function (6.5)
$$Tr(\rho)=({4A^2\over H^2}({F^2\over 4}-BD))^{-1/2}.\eqno(6.14)$$
This quantity is time-independent which can be verified by constructing an
equation satisfied by the quantity $(F^2/4-BD)/H^2$. Combining Eqs.
(6.7)-(6.9) we get
$${d\over dt}({F^2/4-BD\over H^2})=2[2\lambda-(D_1^*-\mu){B\over H}-
(D_1-\mu){D\over H}-(D_2+\lambda){F\over H}]({F^2/4-BD\over H^2}).$$
We see immediately that the first factor on the right-hand side of this
equation is identical with the right-hand side of Eq. (6.6). Accordingly, we
find
$${d\over dt}(({F^2\over 4}-BD){A^2\over H^2})=0.$$
Since the scaling function $H$ is arbitrary, we simplify the following
equations by the choice
$${F^2\over 4}-BD=-H.\eqno (6.15)$$
Setting $Tr\rho=1$, we obtain from Eqs. (6.14) and (6.15) the normalization
constant $A^2=-H/4$.
As a consequence, we can simplify Eqs. (6.7)-(6.9) by eliminating the function
$H$ from these equations. The resulting three equations are
$${dB\over dt}=-2(\lambda+i\omega)B-\mu F+2(D_1-\mu),$$
$${dD\over dt}=-2(\lambda-i\omega)D-\mu F+2(D_1^*-\mu),\eqno(6.16)$$
$${dF\over dt}=-2\mu(B+D)-2\lambda F-4(D_2+\lambda).$$
These equations imply that the function $D$ is complex conjugate to $B$,
provided that the function $F$ is real.

In order to integrate the equations for the time-dependent functions $B,C,D,E$
and $F$ we start with Eqs. (6.12) and (6.13). These equations imply that the
function $E$ is complex conjugate to the function $C$. By solving the coupled
equations we find:
$$C(t)=E^*(t)=u(t)C(0)-v(t)C^*(0),\eqno(6.17)$$
where $u(t)$ and $v(t)$ are given by (4.13) and (4.14) for the two considered
cases: overdamped and underdamped, respectively.
For integrating the system (6.16) we proceed in the same way as for integrating
the system (3.13). With the assumption that $F$ is real and
$$D(t)=B^*(t)=R(t)+iI(t),$$
we obtain explicitly:
$$R(t)={1\over 2}(e^{-2\mu_+t}+e^{-2\mu_-t})\widetilde R+{1\over 2}(e^{-2\mu_+
t}-
e^{-2\mu_-t})({\omega\over\gamma}\widetilde I+{\mu\over 2\gamma}\widetilde F)+
R(\infty),$$
$$I(t)=e^{-2\lambda t}({\mu^2\over\gamma^2}\widetilde I+{\omega\mu\over 2
\gamma^2}
\widetilde F)-$$
$$-{1\over 2}(e^{-2\mu_+t}+e^{-2\mu_-t})({\omega^2\over\gamma^2}\widetilde I+
{\omega\mu\over 2\gamma^2}\widetilde F)-{\omega\over 2\gamma}(e^{-2\mu_+t}-e^{
-2\mu
_-t})\widetilde R+I(\infty),$$
$$F(t)=-e^{-2\lambda t}({2\omega\mu\over\gamma^2}\widetilde I+{\omega^2\over
\gamma
^2}\widetilde F)+$$
$$+(e^{-2\mu_+t}+e^{-2\mu_-t})({\omega\mu\over\gamma^2}\widetilde I+{\mu^2
\over 2\gamma^2}\widetilde F)+{\mu\over\gamma}(e^{-2\mu_+t}-e^{-2\mu_-t})
\widetilde R+
F(\infty),$$
where we used the notations:
$$\mu_\pm=\lambda\pm\gamma, \gamma\equiv\sqrt{\mu^2-\omega^2},$$
$$\widetilde R=R(0)-R(\infty), \widetilde I=I(0)-I(\infty), \widetilde F=F(0)-
F(\infty).$$
We can also obtain the connection between the asymptotic values of $B(t),D(t),
F(t)$ and the coefficients $D_1, D_2, \mu$ and $\lambda$:
$$R(\infty)=ReD(\infty)={\lambda(ReD_1-\mu)+\omega ImD_1+\mu(D_2+\lambda)
\over\lambda^2-\gamma^2},$$
$$I(\infty)=ImD(\infty)={\omega\lambda(ReD_1-\mu)+(\mu^2-\lambda^2)ImD_1+
\omega\mu(D_2+\lambda)\over\lambda(\lambda^2-\gamma^2)},$$
$$F(\infty)=-2{\mu[\lambda(ReD_1-\mu)+\omega ImD_1]+(\lambda^2+\omega^2)(D_2
+\lambda)\over\lambda(\lambda^2-\gamma^2)}.$$
When all explicit expressions for $A,B,C,D,E,F$ and $H$ are introduced into
Eq. (6.5), we obtain an analytical form of the generating function $G(x,y,t)$
which allows us to evaluate the density matrix.

If the constants involved in the generating function satisfy the relations
$$C(0)=0, R(0)=R(\infty), I(0)=I(\infty), F(0)=F(\infty),$$
we obtain the stationary solution
$$C(t)=E(t)=0, R(t)=R(0), I(t)=I(0), F(t)=F(0),$$
so that
$$D(t)=B^*(t)=R(0)+iI(0),$$
$$H(t)=-4A^2(t)=R^2(0)+I^2(0)-F^2(0)/4.$$
Then the stationary solution of Eq. (6.4) is
$$G(x,y,t)={1\over A}\exp\{(1-{F\over H})xy-(Bx^2+B^*y^2)/H\}.\eqno(6.18)$$
In addition, for a thermal bath [17] with
$${m\omega D_{qq}\over \hbar}={D_{pp}\over\hbar m\omega}, D_{pq}=0, \mu=0,$$
the stationary generating function is simply given by
$$G(x,y)={2\lambda\over D_2+\lambda}\exp[{D_2-\lambda\over D_2+\lambda}xy].$$
The same generating function can be found for large times, if the asymptotic
state is a Gibbs state with $\mu=0$. In this case we obtain with Eq. (3.26) and
$\mu=0$
$$D_2=\lambda\coth{\hbar\omega\over 2kT}$$
and $$G(x,y)=(1-\exp(-{\hbar\omega\over kT}))\exp(\exp(-{\hbar\omega\over kT})
xy).$$
The density matrix can be calculated with Eq. (6.3) and yields the
Bose-Einstein distribution
$$\rho=(1-\exp(-{\hbar\omega\over kT}))\exp(-{n\hbar\omega\over kT})\delta_
{nm}.$$

A formula for the density matrix can be written down by applying the relation
(6.3) to the generating function (6.5). We get
$$<m\vert\rho(t)\vert n>={\sqrt{m!n!}\over A}\exp[-(BC^2+DE^2+FCE)/H]\times$$
$$\sum_{n_1,n_2,n_3=0}{(1-{F\over H})^{n_3}(-{B\over H})^{n_1}(-{D\over H})^
{n_2}(2{BC\over H}+{FE \over H})^{m-2n_1-n_3}(2{DE \over
H}+{FC \over H})^{n-2n_2 -n_3}\over
n_1!n_2!n_3!(m-2n_1-n_3)!(n-2n_2-n_3)!}.\eqno(6.19)$$ In
the case that the functions $C(t)$ and $E(t)$ vanish, the
generating function has the form of Eq. (6.18). Then the
elements of the density matrix with an odd sum $m+n$ are
zero: $\rho_{mn}=0$ for $m+n=2k+1$ with $k=0,1,2,...$ The
lowest non-vanishing elements are given with
$\rho_{mn}=\rho_{nm}$ as
$$\rho_{00}={1\over A}, \rho_{20}=-{\sqrt 2 B\over AH}, \rho_{11}={1\over A}
(1-{F\over H}),$$
$$\rho_{22}={2BB^*\over AH^2}+{1\over A}(-{F\over H})^2, \rho_{31}=-(1-{F
\over H}){\sqrt 6B\over AH},
 \rho_{40}={\sqrt 6B^2\over AH^2}.$$
It is also possible to choose the constants in such a way that the functions
$B$ and $D$ vanish at time $t=0$ and $F(0)=H(0)$. Then the density matrix
(6.19) becomes at $t=0 (E=C^*)$:
$$<m\vert\rho(0)\vert n>={1\over\sqrt{m!n!}}(C^*(0))^m(C(0))^n\exp(-\vert C(0)
\vert^2).\eqno(6.20)$$
This is the initial Glauber
packet. The diagonal matrix elements of Eq. (6.20) represent a Poisson
distribution used also in the study of multi-phonon excitations in nuclear
physics. In the particular case when we assume
$$D_1=\mu=0, D_2=\lambda,$$
$$B(0)=D(0)=0, F(0)=H(0)=-4,$$
the differential equations (6.16) yield
$B(t)=D(t)=0$ and $F(t)=H(t)=-4$.
Then the density matrix subject to the initial Glauber packet is (see also
[39])
$$<m\vert\rho(t)\vert n>={1\over\sqrt{m!n!}}(C^*(t))^m(C(t))^n\exp(-\vert C(t)
\vert^2),$$
where $C(t)$ is given by Eq. (6.17).

\vskip 1truecm

{\bf 4. Conclusions}

The Lindblad theory provides a selfconsistent treatment of damping as a
possible
extension of quantum mechanics to open systems. In the present paper first
we studied the damped quantum oscillator by using the Schr\"odinger and
Heisenberg representations. According to this theory we have calculated the
damping of the expectation values of coordinate and momentum and the variances
as functions of time. The resulting time dependence of the expectation values
yields an exponential damping.
Second we have also shown how the quasiprobability distributions can be used to solve
the problem of dissipation for the harmonic oscillator. From the master
equation of the damped quantum oscillator we have derived the corresponding
Fokker-Planck equations in the Glauber $P$, the antinormal ordering $Q$ and
the Wigner $W$ representations and have made a comparative study of these
quasiprobability distributions. We have proven that the variances found from
the Fokker-Planck equations in these representations are the same. We have
solved these equations in the steady state and showed that the Glauber $P$
function (when it exists), the $Q$ and the Wigner $W$ functions are
two-dimensional Gaussians with different widths.
Finally, we have calculated the time evolution of the density matrix. For this
purpose we applied the method of the generating function of the density
matrix. In this case the density matrix can be obtained by taking partial
derivatives of
the generating  function. The generating function depends on a set of
time-dependent coefficients which can be calculated as solutions of linear
differential equations of first order. Depending on the initial conditions
for these coefficients, the density matrix evolves differently in time. For a
thermal bath, when the asymptotic state is a Gibbs state, a Bose-Einstein
distribution results as density matrix. Also for the case that the initial
density matrix is chosen as a Glauber packet, a simple analytical expression
for the density matrix has been derived.
The density matrix can be used in various physical applications where a
Bosonic degree of freedom moving in a harmonic oscillator potential is
damped. For example, one needs to determine nondiagonal transition elements
of the density matrix, if an oscillator is perturbed by a weak electromagnetic
field in addition to its coupling to a heat bath. The density matrix can also
be derived from the solution of the Fokker-Planck equation for the coherent
state  representation.

\vskip 1truecm

{\bf References}

\item{1.}
E. B. Davies, Quantum Theory of Open Systems, Academic Press, 1976

\item{2.}
K. H. Li, Phys. Rep. {\bf 134} (1986) 1

\item{3.}
J. Messer, Acta Phys. Austriaca {\bf 58} (1979) 75

\item{4.}
H. Dekker, Phys. Rep. {\bf 80} (1981) 1

\item{5.}
R.Haake, Springer Tracts in Mod. Phys. {\bf 66} (1973) 98

\item{6.}
V. Gorini, A. Kossakovski, J. Math. Phys. {\bf 17} (1976) 1298

\item{7.}
R. S. Ingarden, A. Kossakowski, Ann. Phys. (N.Y.) {\bf 89} (1975) 451

\item{8.}
V. Gorini, A. Frigerio, M. Verri, A. Kossakowski, E. C. G. Sudarshan,
Rep. Math. Phys. {\bf 13} (1978) 149

\item{9.}
R. S. Ingarden, Acta Phys. Polonica A {\bf 43} (1973) 1

\item{10.}
G. Lindblad, Commun. Math. Phys. {\bf 48} (1976) 119

\item{11.}
A. Kossakowski, Rep. Math. Phys. {\bf 3} (1972) 247

\item{12.}
A. Kossakowski, Bull. Acad. Polon. Sci. Math. Astron. Phys. {\bf 20} (1972)
1021

\item{13.}
G. C. Emch, Algebraic Methods in Statistical Mechanics and Quantum
Field Theory, Wiley, 1972

\item{14.}
V. Gorini, A. Kossakowski, E. C. G. Sudarshan, J. Math. Phys. {\bf 17} (1976)
821

\item{15.}
G. Lindblad, Rep. Math. Phys. {\bf 10} (1976) 393

\item{16.}
P. Talkner, Ann. Phys. (N.Y.) {\bf 167} (1986) 390

\item{17.}
A. Sandulescu, H. Scutaru, Ann. Phys. (N.Y.) {\bf 173} (1987) 277

\item{18.}
A. Pop, A. Sandulescu, H. Scutaru, W. Greiner, Z. Phys. A-Atomic Nuclei {\bf
329} (1988) 357

\item{19.}
A. Isar, A. Sandulescu, W. Scheid, J. Phys. G-Nucl. Part. Phys. {\bf 17} (1991)
385

\item{20.}
A. Isar, A. Sandulescu, Rev. Roum. Phys. {\bf 34} (1989) 1213 (Contribution at
the International School on Nuclear Physics, Poiana Brasov, 1988)

\item{21.}
A. Isar, A. Sandulescu, W. Scheid (to be published)

\item{22.}
A. Isar, A. Sandulescu, W. Scheid, J. Math. Phys. {\bf 32} (1991) 2128

\item{23.}
H. Dekker, M. C. Valsakumar, Phys. Lett. A {\bf 104} (1984) 67

\item{24.}
H. Dekker, Phys. Lett. A {\bf 74} (1979) 15

\item{25.}
H. Dekker, Phys. Rev. A {\bf 16} (1979) 2126

\item{26.}
H. Dekker, Phys. Lett. A {\bf 80} (1980) 369

\item{27.}
H. Hofmann, C. Gr\'egoire, R. Lucas, C. Ng\^o, Z. Phys. A-Atomic Nuclei
{\bf 293} (1979) 229

\item{28.}
H. Hofmann, P. J. Siemens, Nucl. Phys. A {\bf 275} (1977) 464

\item{29.}
R. W. Hasse, Nucl. Phys. A {\bf 318} (1979) 480

\item{30.}
E. M. Spina, H. A. Weidenm\"uller, Nucl. Phys. A {\bf 425} (1984) 354

\item{31.}
C. W. Gardiner, M. J. Collet, Phys. Rev. A {\bf 31} (1985) 3761

\item{32.}
T. A. B. Kennedy, D. F. Walls, Phys. Rev. A {\bf 37} (1988) 152

\item{33.}
C. M. Savage, D. F. Walls, Phys. Rev. A {\bf 32} (1985) 2316

\item{34.}
G. S. Agarwal, Phys. Rev. {\bf 178} (1969) 2025

\item{35.}
G. S. Agarwal, Phys. Rev. A {\bf 4} (1971) 739

\item{36.}
S. Dattagupta, Phys. Rev. A {\bf 30} (1984) 1525

\item{37.}
N. Lu, S. Y. Zhu, G. S. Agarwal, Phys. Rev. A {\bf 40} (1989) 258

\item{38.}
S. Jang, C. Yannouleas, Nucl. Phys. A {\bf 460} (1986) 201

\item{39.}
S. Jang, Nucl. Phys. A {\bf 499} (1989) 250

\item{40.}
S. Chaturvedi, P. D. Drummond, D. F. Walls, J. Phys. A-Math. Gen. {\bf 10}
(1977) L187

\item{41.}
P. D. Drummond, C. W. Gardiner, J. Phys. A-Math. Gen. {\bf 13} (1980) 2353

\item{42.}
P. D. Drummond, C. W. Gardiner, D. F. Walls, Phys. Rev. A {\bf 24} (1981) 914

\item{43.}
E. P. Wigner, Phys. Rev. {\bf 40} (1932) 749

\item{44.}
E. J. Glauber, Phys. Rev. {\bf 131} (1963) 2766

\item{45.}
E. J. Glauber, Phys. Rev. Lett. {\bf 10} (1963) 84

\item{46.}
E. C. G. Sudarshan, Phys. Rev. Lett. {\bf 10} (1963) 277

\item{47.}
W. Weidlich, H. Risken, H. Haken, Z. Phys. {\bf 204} (1967) 223

\item{48.}
M. Lax, W. H. Louisell, IEEE J. Q. Electron. {\bf 3} (1967) 47

\item{49.}
K. E. Cahill, R. J. Glauber, Phys. Rev. A {\bf 117} (1969) 1882

\item{50.}
R. J. Glauber, in Laser Handbook, ed. by F. T. Arecchi and E. O.
Schultz-Dubois, North-Holland, 1972

\item{51.}
J. R. Klauder, E. C. G. Sudarshan, Fundamentals of Quantum Optics,
Benjamin, 1968

\item{52.}
K. E. Cahill, Phys. Rev. {\bf 180} (1969) 1239

\item{53.}
K. E. Cahill, Phys. Rev. {\bf 180} (1969) 1244

\item{54.}
W. H. Louisell, Quantum Statistical Properties of Radiation, Wiley, 1973

\item{55.}
C. W. Gardiner, Handbook of Stochastic Methods, Springer, 1982

\item{56.}
G. E. Uhlenbeck, L. S. Ornstein, Phys. Rev. {\bf 36} (1930) 823

\item{57.}
M. C. Wang, G. E. Uhlenbeck, Rev. Mod. Phys. {\bf 17} (1945) 323

\item{58.}
M. Hillery, R. F. O 'Connell, M. O. Scully, E. P. Wigner, Phys. Rev. {\bf 106}
(1984) 121

\bye